\documentclass[modern]{aastex631}
\usepackage{graphicx,natbib,color,bm,url,times, hyperref}
\usepackage{amsmath}
\usepackage{amssymb}
\usepackage{euscript}
\usepackage{float}
\usepackage{tabularx}
\usepackage{array}
\usepackage{orcidlink}
\usepackage{hyperref}
\hypersetup{breaklinks=true, colorlinks=true, citecolor=blue}
\usepackage{color}
\usepackage{capt-of}

\newcommand{\beq}{\begin{equation}}
\newcommand{\eeq}{\end{equation}}
\newcommand{\Eq}[1]{Equation~(\ref{#1})}
\newcommand{\Eqs}[2]{Equations~(\ref{#1}) and~(\ref{#2})}
\newcommand{\Eqss}[2]{Equations~(\ref{#1})--(\ref{#2})}
\newcommand{\App}[1]{Appendix~\ref{#1}}
\newcommand{\Sec}[1]{Section~\ref{#1}}
\newcommand{\Secs}[2]{Sections~\ref{#1} and \ref{#2}}
\newcommand{\Fig}[1]{Figure~\ref{#1}}
\newcommand{\Figs}[2]{Figures~\ref{#1} and \ref{#2}}

\newcommand{\bra}[1]{\langle #1\rangle}
\newcommand{\bbra}[1]{\left\langle #1\right\rangle}

\newcommand{\vv}{\mbox{\boldmath $v$}}
\newcommand{\kk}{\mbox{\boldmath $k$}}
\newcommand{\xx}{\mbox{\boldmath $x$}}
\newcommand{\ww}{\mbox{\boldmath $w$}}
\newcommand{\bkap}{\mbox{\boldmath $\kappa$}}
\newcommand{\ea}{\mbox{\boldmath $\hat{a}$}}
\newcommand{\eu}{\mbox{\boldmath $\hat{u}$}}

\newcommand{\ii}{{\rm i}}

\newcommand{\dd}{{\rm d} {}}
\newcommand{\bnabla}{\mbox{\boldmath $\nabla$}}
\newcommand{\cross}{\mbox{\boldmath $\times$}}
\newcommand{\cendot}{\mbox{\boldmath $\cdot$}}

\newcommand{\ra}{{\rm Ra}}
\newcommand{\pr}{{\rm Pr}}

\begin{document}

\title{Turbulent Convection: Modal Equations and Energy Pathways}

\author{S. Sridhar}
\email{ssridhar811@gmail.com}
\affiliation{Inter--University Centre for Astronomy \& Astrophysics, Post Bag 4, Ganeshkhind, Pune 411007, India}

\author{Nishant K. Singh\orcidlink{0000-0001-6097-688X}}
\email{nishant@iucaa.in}
\affiliation{Inter--University Centre for Astronomy \& Astrophysics, Post Bag 4, Ganeshkhind, Pune 411007, India}

\begin{abstract}
We present a framework for studying high Rayleigh number turbulent convection to better understand stellar and planetary convection zones. Utilizing the statistical symmetries of the fully developed turbulent state of Boussinesq convection, we identify relevant mean and fluctuating quantities. After validating these symmetry assumptions through numerical simulations, we formulate the governing equations. Vertical profiles of key physical
quantities in the saturated turbulent state are explored in the simulations. To develop a modal theory, we use Fourier expansions, review linear theory, and use the Craya-Herring velocity decomposition. The modal equations we derive describe high Rayleigh-number turbulent convection dynamics self-consistently in terms of nonlinear interactions between three mode types: growing gravity modes, decaying gravity modes, and horizontal modes. Energy extracted by the growing modes from the superadiabatic background subsequently follows multiple pathways toward dissipation, enabled by the mode couplings. Among these, the traditionally dominant pathway is the turbulent cascade of the growing modes themselves. Reduced modal equations capture this pathway, precisely describing (i) mutual interactions between growing modes, and (ii) the excitation of decaying and horizontal modes, which are subordinate to the growing modes. Determining the relative efficiency of the pathways requires investigating their modal spectra using numerical simulations and kinetic models.
\end{abstract}

%\maketitle

\section{Introduction}
Convection in stars and planets transports heat and plays an important role in their structure and appearance; see the reviews \citet{ss20,gar21} and references therein. The fluid flows are turbulent in many cases and typically extend over several pressure scale heights, making theoretical modelling challenging. Recent helioseismic measurements have revealed very weak convective flow amplitudes at large scales in the Sun, which are at odds with the standard mixing length theory of stellar convection   
\citep{hds12,hgs16,pro21,bir24}. This discrepancy, often referred to as ``the solar convective conundrum'', provided the motivation for modified models of stellar convection 
\citep[see, e.g.,][]{b16,fh16,vas21,kap21,kap25}. The time seems opportune to revisit the foundations of the theory of convection.

It is common to consider analogous phenomena in a simpler context, Rayleigh-Benard convection (RBC), which occurs in a fluid layer sandwiched between two horizontal plates,
heated from below and cooled from above \citep[see, e.g.,][]{kad01,ver18}.
The system is often described theoretically in the Boussinesq approximation (constant density, incompressible velocity field and temperature fluctuations resulting in buoyant forcing). RBC has been extensively studied experimentally, theoretically and through numerical simulations; see, e.g., \citet{sig94, agl09, ver18, ls24}, and references therein. The structure and dynamics of RBC have been explored for a variety of control parameters---such as the Rayleigh number, thermal Prandtl number, and aspect ratios of the flow domain---and boundary conditions \citep{htb03,jd09,pss18,vie24,alam25}.
Much of the work is related to phenomena relevant to fluid confined by thermally insulating and impenetrable side walls. A somewhat surprising experimental fact, noted by \citet{sig94}, is the occurrence of a persistent mean flow at high Rayleigh numbers. This large-scale circulation affects the entire structure of the convecting fluid, and phenomenological theories have relied on distinguishing the bulk of the fluid from the boundary layers.

However, mean flows need not occur in the laterally unconfined geometry of a plane-parallel approximation of a non-rotating star or planet. Although large-scale circulation may be present in high Rayleigh number simulations with periodic lateral boundary conditions, the corresponding mean flow is found to be much weaker \citep{htb03}.
\citet{lss61,yam63} considered turbulent RBC with no side walls or mean flow; henceforth, we refer to this setup as Boussinesq convection, to distinguish it from RBC with side walls and mean flows. Our aim is to present a modal framework for studying fully developed turbulent convection that treats the system in its entirety, without the aforementioned separation into bulk and boundary layers. We exploit the natural statistical symmetries---statistical homogeneity and isotropy in the horizontal directions---that emerge in the absence of side walls and mean flow, and formulate the problem allowing for general vertical variations of horizontally-averaged (or `mean') quantities. Hence, the resulting modal equations discussed in \Sec{sec:mod-path} provide the means for a treatment of anisotropic turbulent convection, in which the mean velocities vanish, and the mean temperature profile and convective heat fluxes are determined self-consistently along with the modal spectra.

In \Sec{sec:define} we state the Boussinesq equations and boundary conditions governing the system, discuss the natural symmetries, and define the mean and fluctuating velocity and temperature fields. These symmetry assumptions are then validated through numerical simulations\footnote{Because the lateral extent is necessarily finite in numerical simulations, periodic boundary conditions are appropriate in the horizontal directions.}, and used in \Sec{sec:gov} to derive the equations governing the evolutions of mean and fluctuating quantities. Various convective fluxes are identified, and the vertical profiles of several key physical quantities, in the saturated state of turbulence, are explored through numerical simulation. In \Sec{sec:fou-lin}, the boundary conditions are satisfied through Fourier expansion, and the linear theory of modes is briefly reviewed.

The nonlinear development is taken up in \Sec{sec:non-dev} through systematic use of the Fourier expansion. The mean temperature vertical profile is solved for in terms of the means of temperature-velocity fluctuations determining the enthalpy flux vertical profile. The governing equations describe the self-consistent evolution of the velocity and temperature fluctuations. Then, using a Craya-Herring decomposition of the velocity field, the equations are cast in terms of a uniform set of variables, instead on a mixture of vector velocity and temperature scalar fields. The final step is taken in \Sec{sec:mod-path}, in the passage to the modal variables which also happen to reduce to eigenmodes in the linear limit. These modal equations describe all possible combinations of quadratic interactions between three mode types and hence allow for multiple turbulent pathways for energy to dissipate. Highly reduced modal equations describe the case of the traditionally dominant pathway considered by \citet{lss61}. We conclude in \Sec{sec:con} with a discussion of applications of the reduced equations to derive kinetic models, and the exploration of modal power-and cross-spectra in conjunction with numerical simulations. 

\section{Describing Turbulent Convection}\label{sec:define}
\subsection{The Boussinesq Equations}\label{sec:bous}
We consider thermal convection in which a fluid, contained between
two horizontal planes separated by a distance $d$, is heated from
below and cooled from above. The unperturbed system is in
hydrostatic equilibrium under a constant downward gravitational
acceleration $-g\hat{z}$, with a linearly decreasing vertical
temperature profile, $T_0(z) = T_{\rm{bot}} - \beta_0 z$.
In the perturbed system the fluid moves with velocity
$\vv(\xx, t) = \vv(x,y,z,t)$, resulting in motion-induced temperature
variations $T'(\xx, t)$---over and above the linear profile---and
pressure variations $p'(\xx, t)$---relative to the hydrostatic
pressure. The equations governing the time evolution of $\vv$
and $T'$ are the incompressible Navier-Stokes equations including
buoyancy, combined with the entropy equation under the Boussinesq
approximation \citep{sv60, tri88}:

\begin{subequations}
\begin{align}
\bnabla\cendot\,\vv &\;=\; 0\,,
\label{eq:incom}\\[1ex]
\frac{\partial \vv}{\partial t} \,+\, \left(\vv\,\cendot\bnabla\right)\vv
&\;=\; -\frac{\bnabla p'}{\rho} \,+\, g\alpha T'\hat{z}
\,+\, \nu\bnabla^2 \vv\,, 
\label{eq:ns}\\[1ex]
\frac{\partial T'}{\partial t} \,+\, \left(\vv\,\cendot\bnabla\right)T'
&\;=\; \beta v_z \,+\, \chi\nabla^2 T'\,.
\label{eq:ent}
\end{align}
\noindent 
\mbox{\Eq{eq:incom} requires $p'(\xx, t)$ to obey Poisson's equation:}
\beq
\frac{\nabla^2 p'}{\rho}
\;=\; -\bnabla\cendot\left[(\vv\,\cendot\bnabla)\vv\right]
\,+\, g\alpha\frac{\partial T'}{\partial z}\,.
\label{eq:pres}
\eeq
\end{subequations}
\noindent
The density $\rho$, coefficient of thermal expansion $\alpha$, kinematic viscosity $\nu$, and thermal diffusivity $\chi$ are given constant fluid properties. The superadiabatic temperature gradient $\beta = \beta_0 - \beta_{\rm ad}\,$, where $\beta_{\rm ad}$ is the magnitude of the adiabatic temperature gradient, is an important control parameter. 

\Eqss{eq:incom}{eq:pres} should be solved with suitable boundary conditions on 
$\left(\vv, T',p'\right)$. For vertical boundary conditions we follow \citet{lss61,yam63} by using the so-called free boundary conditions on the bottom ($z=0$) and top ($z=d$) planes: 
\beq
\frac{\partial v_x}{\partial z} \,=\, \frac{\partial v_y}{\partial z} \,=\, 0\,;\;
v_z \,=\, 0\,;\;
T' \,=\, 0\,;\;
\frac{\partial p'}{\partial z} \,=\, 0\,.
\label{z-bc}
\eeq
\noindent
The fluid freely slips tangentially along the bottom and top boundaries but neither penetrates them nor exerts tangential stress on them. For applications to planetary and stellar convection layers, the fluid may be thought of as infinitely extended in the horizontal directions; the appropriate lateral boundary conditions are then that 
$\left(\vv, T',p'\right)$ vanish as $x$ and $y$ go to infinity. Numerical simulations often confine the fluid within a square box of size $L\times L\times d$ and impose periodic boundary conditions on $\left(\vv, T',p'\right)$ in the $x$ and $y$ directions. In this section, we develop a theory applicable to both infinite and periodic lateral boundary conditions.

\subsection{Symmetries, Horizontal Averages, and Fluctuations}\label{sec:sym}
The Boussinesq system of the previous section has a well-known \emph{reflection symmetry} about the mid-plane. Let $z' = z - d/2$ be the distance from the mid-plane. Then \Eqss{eq:incom}{z-bc} and the lateral boundary conditions are invariant under the combined transformation: 
\beq
\{x,y,z'\} \to \{x,y, -z'\}\,;\quad
\{v_x,v_y,v_z\} \to \{v_x,v_y, -v_z\}\,;\quad
T' \to -T'\,;\quad
p' \to p'\,.
\label{eq:dyn-symm}
\eeq
While this is a general symmetry, individual solutions will typically not respect it because of initial conditions. However, the symmetry appears to be restored statistically in fully developed turbulent convection; see \Sec{sec:num2}.
 
We now discuss the statistical symmetries that may be expected to hold in fully developed turbulence. \emph{Statistical homogeneity} in the horizontal directions is a natural symmetry. For the case of infinite lateral boundary conditions, relevant for models of planetary and stellar convective layers, \emph{statistical isotropy} in the horizontal directions is also a natural symmetry. While isotropy is not a symmetry for numerical simulations in a square box using periodic lateral boundary conditions, statistical symmetry under the \emph{reflection} ($v_x \to -v_x$ or $v_y \to -v_y$) holds. We do not invoke stationarity in time, because we want to allow for a description that includes the time-dependent approach to the saturation of turbulence.

Homogeneity implies that horizontal averaging of dynamical variables is physically meaningful. If $q(\xx, t)$ represents any of $\left(\vv, T',p'\right)$ or functions of these quantities, the mean $\overline{q}$ is defined as: 
\beq
\overline{q}(z, t) \;=\; \frac{1}{L^2}\int_{-L/2}^{L/2}\int_{-L/2}^{L/2} q(\xx, t)\,\dd x\,\dd y\,.
\label{q-av}
\eeq
\noindent
This definition applies to periodic lateral boundary conditions, taking the limit $L\to\infty$ for the infinite case. For both cases we have,  
\beq
\overline{\frac{\partial q}{\partial x}} \,=\, \overline{\frac{\partial q}{\partial y}} \,=\, 0\,;\; \overline{\frac{\partial q}{\partial z}} \,=\, \frac{\partial \overline{q}}{\partial z}\,.
\label{q-der-av}
\eeq
Horizontally averaging \Eq{eq:incom} and using \Eq{q-der-av}, we obtain $\partial \overline{v}_z/\partial z = 0\,$, so $\overline{v}_z$ must be independent of $z$. Since \Eq{z-bc} requires $v_z = 0$ at $z=0$ and $z=d$, we must have $\overline{v}_z = 0$ throughout the fluid. Homogeneity, by itself, does not require that either 
$\overline{v}_x$ or $\overline{v}_y$ vanish anywhere. However, isotropy or reflection symmetry suffice to require that $\overline{v}_x = \overline{v}_y = 0\,$, so the 
mean velocity vanishes in both cases: 
\beq
\overline{\vv} \,=\, \mathbf{0}\,.
\label{vel-av}
\eeq
\noindent
Since $\overline{\vv} = 0\,$, the velocity $\vv(\xx, t)$ is a zero-mean fluctuating vector field. However, the means of temperature and pressure, $\overline{T'}(z, t)$ and 
$\overline{p'}(z, t)\,$, do not vanish. So we decompose the temperature and pressure into mean and fluctuating components:
\beq
T'(\xx, t) \,=\, \overline{T'}(z, t) \,+\, \theta(\xx, t)\,;\quad 
p'(\xx, t) \,=\, \overline{p'}(z, t) \,+\, p(\xx, t)\,,
\label{tp-decomp}
\eeq
\noindent
where $\theta$ and $p$ are zero-mean fluctuating scalar fields (i.e., $\overline{\theta} = 0, \,\overline{p} = 0$).

Homogeneity and isotropy (or reflection symmetry) also impose strong constraints on the means of products of $\left(\vv, \theta, p\right)$ taken at multiple spatial locations. In this paper we only require single-point means of quadratic quantities involving 
$\vv$ and $\theta$. Of these, the off-diagonal components of the Reynolds stress 
$\overline{v_i v_j}$, as well as $\,\overline{\theta v_x}$, and $\overline{\theta v_y}$, vanish:
\beq
\overline{v_x v_y} \,=\, 0\,;\quad
\overline{v_x v_z} \,=\, 0\,;\quad
\overline{v_y v_z} \,=\, 0\,;\quad
\overline{\theta v_x} \,=\, 0\,;\quad
\overline{\theta v_y} \,=\, 0\,;
\label{quad-av-zero}
\eeq
\noindent
In \Sec{sec:num1} we test these statistical expectations in \Eqs{vel-av}{quad-av-zero} through numerical simulations. The relevant non-vanishing means are:
\beq
\overline{T'}(z, t)\,;\quad
\overline{v_x^2}(z, t)\,=\, \overline{v_y^2}(z, t)\,;\quad
\overline{v_z^2}(z, t)\,;\quad 
\overline{\theta^2}(z, t)\,;\quad
\overline{\theta v_z}(z, t)\,.
\label{quad-av-non}
\eeq
\noindent
The saturated vertical profiles of these quantities is explored in \Sec{sec:num2}.

\subsection{Numerical Validation of Symmetry Assumptions}\label{sec:num1}
The Boussinesq \Eqss{eq:incom}{eq:ent} are solved numerically by casting them in the standard dimensionless form \citep{dr04}. Measuring $(x, y, z)$ in units of $d$, $\,t$ in units of $(d^2/\chi)$, $\,(v_x, v_y, v_z)$ in units of $(\chi/d)$, $\,T'$ in units of $(\beta d)$, and $\,(p'/\rho)$ in units of $(\chi/d)^2\,$, the dimensionless Boussinesq equations
are:\footnote{To avoid cumbersome notation, we retain the original
variables for dimensionless quantities. The context---specifically
the variable values cited in the figures---will distinguish them
from their dimensional counterparts.}

\begin{subequations}
\begin{align}
\bnabla\cendot\,\vv &\;=\; 0\,,
\nonumber\\[1ex]
\frac{\partial \vv}{\partial t} \,+\, \left(\vv\,\cendot\bnabla\right)\vv
&\;=\; -\frac{\bnabla p'}{\rho} \,+\, \ra .\pr\,T'\hat{z}
\,+\, \pr\,\bnabla^2 \vv\,, 
\nonumber\\[1ex]
\frac{\partial T'}{\partial t} \,+\, \left(\vv\,\cendot\bnabla\right)T'
&\;=\; v_z \,+\, \nabla^2 T'\,,
\nonumber
\end{align}
\end{subequations}
where the Rayleigh number $\ra$ and Prandtl number $\pr$,  
\beq
\ra \;=\; \frac{g\alpha\beta d^4}{\chi\nu}\,,\qquad \pr \;=\; \frac{\nu}{\chi}\,,
\label{eq:rapr}
\eeq
are important dimensionless parameters of the system. $p'(\xx, t)$ is chosen to enforce incompressibility. The boundary conditions on all variables remain unchanged.  

We solved the Boussinesq equations in dimensionless form using the \textsc{Pencil Code}\footnote{\href{https://pencil-code.nordita.org/}{https://pencil-code.nordita.org/}} \citep{pen21}, a publicly available, high-order finite-difference code extensively used for problems involving convection and turbulence. 

We ran a number of simulations for a range of values of $\ra$ and $\pr$,
in square-box domains of different aspect ratios. After obtaining
convergent and consistent measurements in the simulations,
we present the results of one simulation
which was performed at the highest $\ra$, in the flattest box,
and at the finest resolution we could afford. 
The domain was a square box of size $8\times 8 \times 1$ in dimensionless units. Based on convergence tests performed in smaller domains, a grid of $640\times 640 \times 160$ points was chosen. The variables were subjected to the free vertical boundary conditions of \Eq{z-bc} and periodic boundary conditions in the lateral directions. The code computes pressure variations using the projection method \citep{cho67} to ensure flow incompressibility. We performed this simulation at $\ra=10^6$ and $\pr=1$, initializing with $\vv = \mathbf{0}$ and weak, zero-mean Gaussian fluctuations in the temperature field with $\theta_{\rm rms}=10^{-10}$.

\begin{figure}[hbt!]
\begin{center}
\includegraphics[width=\columnwidth]{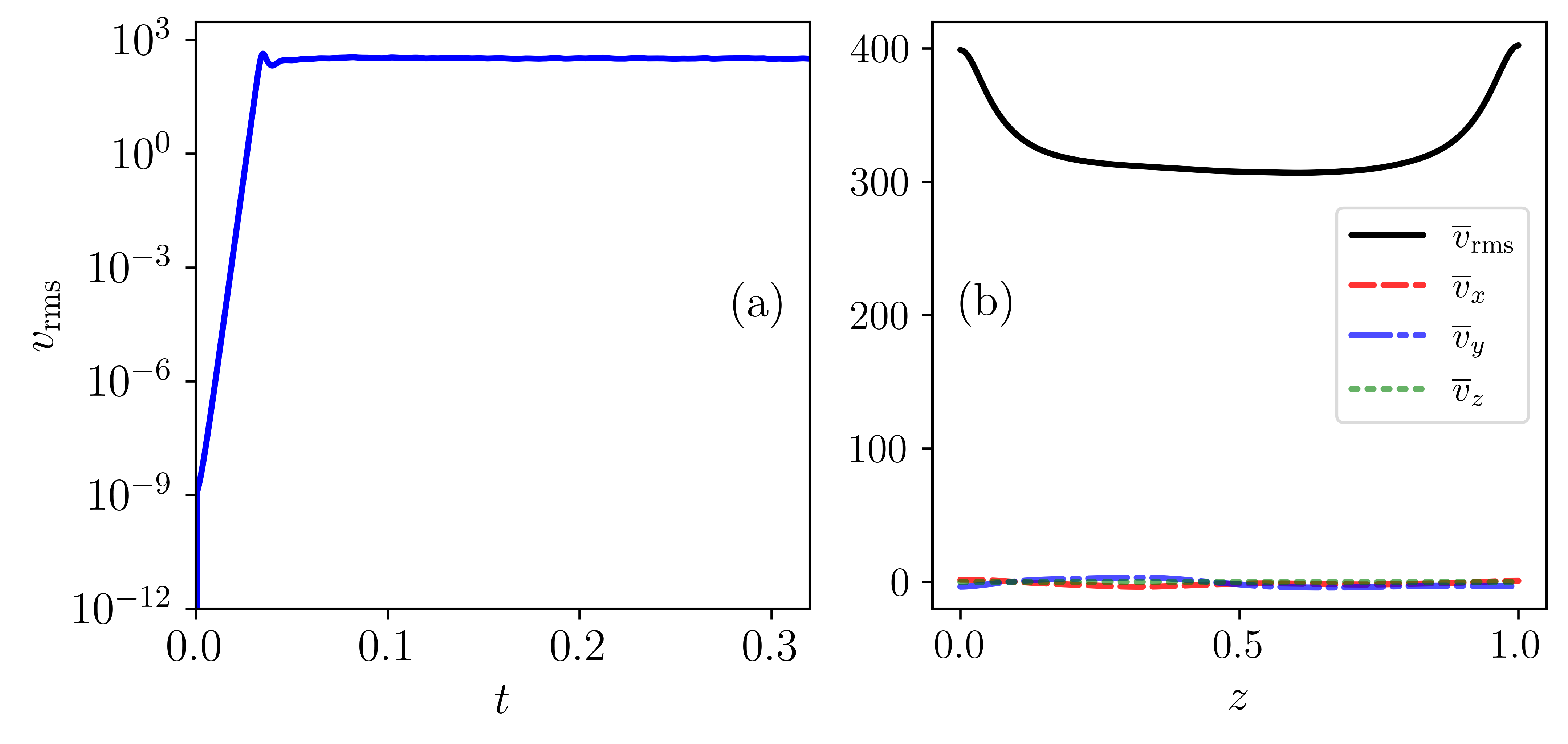}
\end{center}
\caption[]{(a) Exponential growth of $v_{\rm rms}$ in the linear regime, until it reaches a saturated state. (b) Vertical profiles of $\overline{v}_{\rm rms}$, 
$\overline{v}_x$, $\overline{v}_y$, and $\overline{v}_z$.}
\label{urms_t}
\end{figure}

\begin{figure}[hbt!]
\begin{center}
\includegraphics[width=\columnwidth]{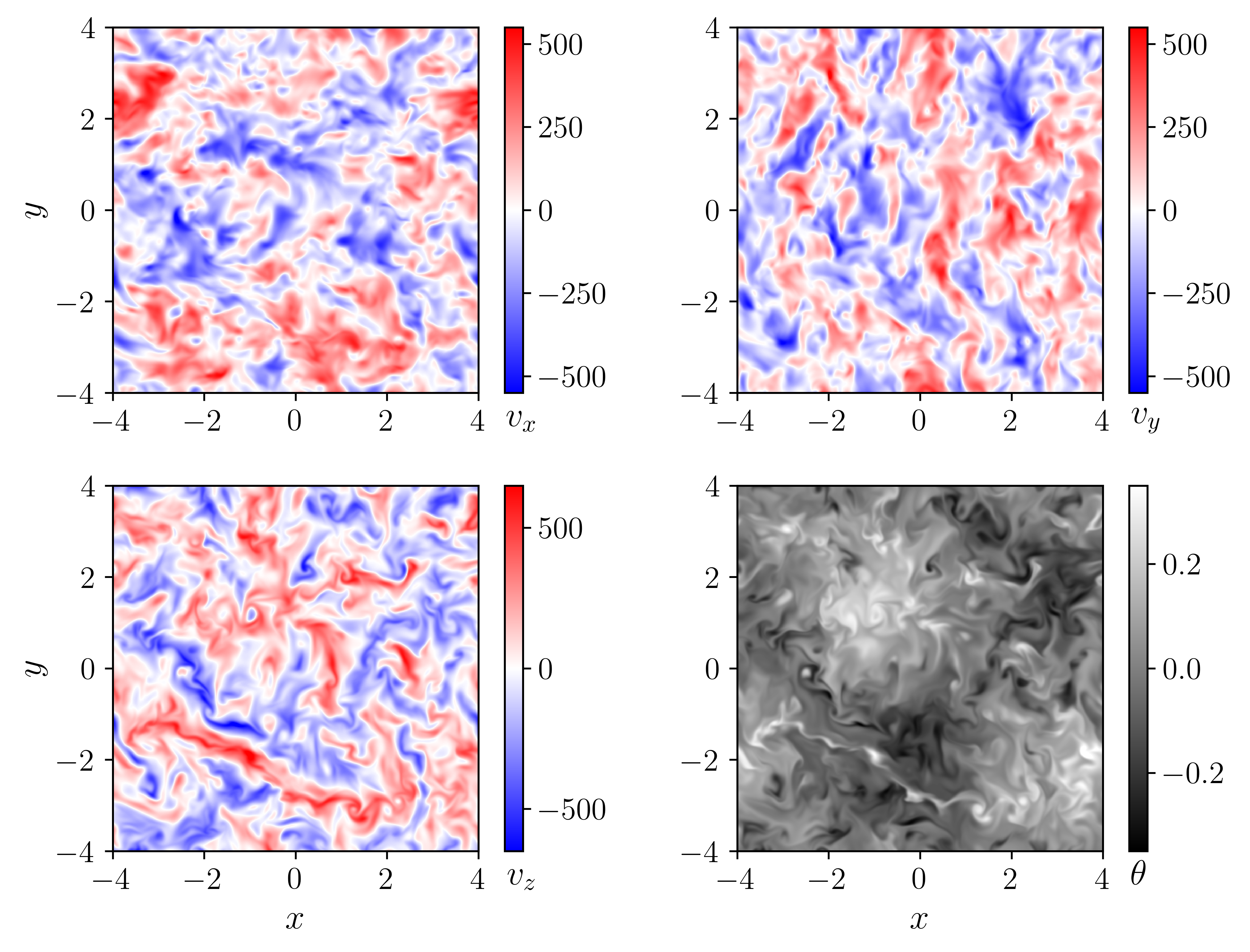}
\end{center}
\caption[]{Snapshot of velocity and temperature variations in the mid-plane $(z=0.5)$.}
\label{flow_theta_midz}
\end{figure}

Panel~(a) of \Fig{urms_t} shows a log-linear plot of the box-averaged root-mean-squared velocity $v_{\rm rms}$ versus time. In the linearly unstable regime, $v_{\rm rms}$ grows exponentially from very small values before entering a brief nonlinear phase. It then saturates at a value $\sim 325$, after which it remains nearly constant. Henceforth, all figures display physical quantities in the saturated state of turbulent convection.

Since $v_{\rm rms} \simeq 325$, the eddy turnover time is shorter than the vertical diffusion time by the same factor $\simeq 325$. Because spatially averaged quantities typically exhibit fluctuations on the scale of the eddy turnover time, their long-term behavior is better captured by averaging over several such periods. 

Panel~(b) of \Fig{urms_t} displays the vertical profiles of various horizontally averaged velocities in the saturated state. The solid black curve shows $\overline{v}_{\rm rms}$, the root-mean-squared velocity at a fixed $z$; it takes its maximum value of about $400$ near the top/bottom boundaries, dropping to a nearly constant value of about $300$ in the bulk of the fluid. Also shown in Panel~(b) are $\left(\overline{v}_x, \,\overline{v}_y, \,\overline{v}_z\right)$, all of which are consistent with zero to within an accuracy of about $\pm 1\%$ of $v_{\rm rms}$. 

Even though $\left(\overline{v}_x, \,\overline{v}_y, \,\overline{v}_z\right)$ are all very small, there are significant fluctuations of physical quantities, evident from the snapshots of velocity components $(v_x, v_y, v_z)$ and temperature fluctuation $\theta$ in \Fig{flow_theta_midz} taken at the mid-plane $(z=0.5)$. All four quantities show somewhat random variations in the horizontal directions. As before, the reference scale for velocities can be taken as $v_{\rm rms} \simeq 325$. Also, $\theta_{\rm rms} \simeq 0.13$. With these reference values in mind, we note that the velocity and temperature fluctuations are strong and of order unity. Also, $v_z$ and $\theta$ are highly correlated.

\begin{figure}[hbt!]
\begin{center}
\includegraphics[width=\columnwidth]{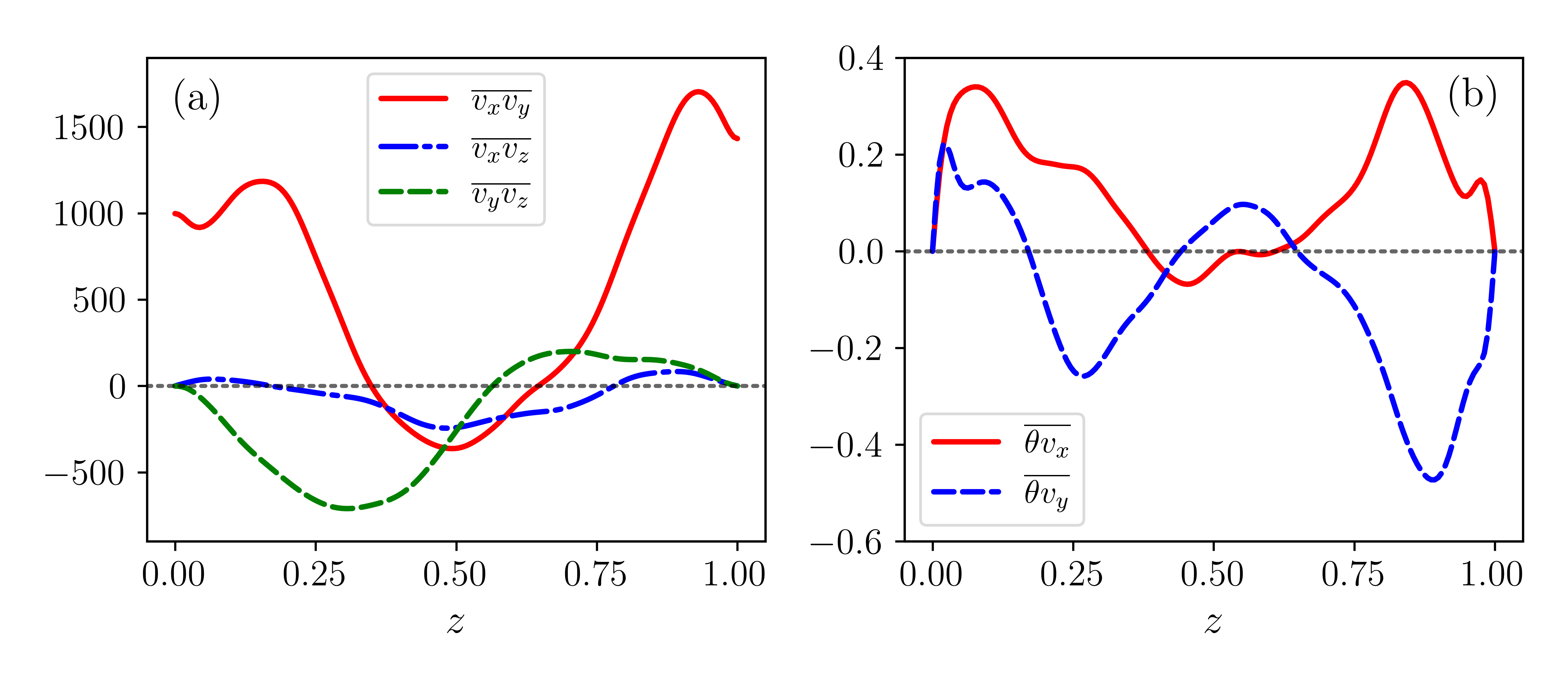}
\end{center}
\caption[]{(a) Vertical profiles of $\overline{v_xv_y},\; \overline{v_xv_z},\; \overline{v_yv_z}$. (b) Vertical profiles of $\overline{\theta v_x},\; \overline{\theta v_y}$.
}\label{2ndmoments}
\end{figure}

Panel~(a) of \Fig{2ndmoments} shows the vertical profiles of the off-diagonal components of the Reynolds stress, $\left(\overline{v_xv_y}, \,\overline{v_xv_z}, \,\overline{v_yv_z}\right)$. Since the dimensionless reference scale for these second-order moments is $v_{\rm rms}^2 \simeq 10^5$, we see that the off-diagonal components are consistent with zero to within an accuracy of about $1\%$ of $v_{\rm rms}^2$. Panel~(b) of \Fig{2ndmoments} shows the vertical profiles of $\overline{\theta v_x}$ and 
$\overline{\theta v_y}\,$. When compared with the dimensionless reference scale for temperature-velocity correlations  $\theta_{\rm rms}v_{\rm rms}/3 \simeq 14$, 
both $\overline{\theta v_x}$ and $\overline{\theta v_y}$ are consistent with zero to within an accuracy of about $3\%$ of $\theta_{\rm rms}v_{\rm rms}/3$.

Therefore, we conclude that the statistical symmetry assumptions in \Eqs{vel-av}{quad-av-zero} have been verified, at least in the saturated state, by the Boussinesq simulations.

\section{Governing Equations and Vertical Profiles}\label{sec:gov}
\subsection{Governing Equations}\label{sec:dyn-eqns}
The mean dynamical equations are obtained by horizontally averaging the Boussinesq equations:  
\begin{subequations}
\begin{align}
&\frac{\partial\, \overline{v_z^2}}{\partial z} \;=\; 
-\frac{\,1\,}{\rho}\frac{\partial\, \overline{p'}}{\partial z} \,+\, g\alpha \overline{T'}\,, 
\label{eq:ns-av}\\[1ex]
&\frac{\partial\overline{T'}}{\partial t} \,+\, \frac{\partial}{\partial z}\left[\overline{\theta v_z} \,-\, \chi\frac{\partial\overline{T'}}{\partial z}\right] \;=\; 0\,,
\label{eq:ent-av}
\end{align}
\end{subequations}
\noindent
relate $\overline{p'}$ and $\overline{T'}$ to single-point, second-order means of the fluctuations. Averaging the vertical boundary conditions in \Eq{z-bc}, and noting that $\overline{\vv}$ vanishes everywhere, we have at $z=0, d\,$:
\beq
\overline{T'} \;=\; 0\,; \qquad \frac{\partial\overline{p'}}{\partial z} \;=\; 0\,. 
\label{eq:bc-tp-av}
\eeq
\noindent
Regarding lateral boundary conditions, either $\overline{T'}$ and $\overline{p'}$ vanish as $x$ and $y$ go to infinity, or they are periodic functions of $x$ and $y$.

Equations governing fluctuating quantities $\left(\vv, p, \theta\right)$ are obtained by subtracting the equations for mean quantities from the full Boussinesq equations:
\begin{subequations}
\begin{align}
\bnabla\cendot\,\vv &\;=\; 0\,,
\label{eq:incom-fl}\\[1ex]
\frac{\partial \vv}{\partial t} \,+\, \left(\vv\,\cendot\bnabla\right)\vv
\,-\, \frac{\partial\, \overline{v_z^2}}{\partial z}\hat{z}  
&\;=\; -\frac{\bnabla p}{\rho} \,+\, g\alpha\theta\hat{z}
\,+\, \nu\bnabla^2 \vv\,, 
\label{eq:ns-fl}\\[1ex]
\frac{\partial \theta}{\partial t} \,+\, \left(\vv\,\cendot\bnabla\right)\theta
\,-\, \frac{\partial}{\partial z}\overline{\theta v_z}
&\;=\; \left(\!\beta - \frac{\partial \overline{T'}}{\partial z}\right)\!v_z \,+\, 
\chi\nabla^2 \theta\,,
\label{eq:ent-fl}
\end{align}
\noindent 
\mbox{where $p(\xx, t)$ obeys the Poisson equation,}
\beq
\frac{\nabla^2 p}{\rho} \;=\; -\bnabla\cendot\left[(\vv\,\cendot\bnabla)\vv\right]
\,+\, \frac{\partial^2}{\partial^2 z}\overline{v_z^2}
\,+\, g\alpha\frac{\partial\theta}{\partial z}\,.
\label{eq:pres-fl}
\eeq
\end{subequations}
\noindent
The vertical boundary conditions at $z=0,d\,$ are: 
\beq
\overline{T'} \,=\, 0\,;\quad \frac{\partial v_x}{\partial z} \,=\, \frac{\partial v_y}{\partial z} \,=\, 0\,;\quad
v_z \,=\, 0\,;\quad
\theta \,=\, 0\,;\quad
\frac{\partial p}{\partial z} \,=\, 0\,.
\label{z-bc-fl}
\eeq
\noindent
The lateral boundary conditions are that either the quantities $\left(\vv, p, \theta\right)$ vanish as $x$ and $y$ go to infinity, or they are periodic functions of $x$ and $y$. The quantities $\left(\overline{T'}, \vv, \theta\right)$ must be chosen at the initial time $t=0$ in accordance with the vertical and lateral boundary conditions and  the incompressibility condition $\bnabla\cendot\,\vv = 0\,$.\footnote{Perhaps the simplest choice is the one we made in the numerical simulations of \Sec{sec:num1}: $\vv(\xx, 0) = {\mbox{\boldmath $0$}}$, while $\theta(\xx, 0)$ is drawn from a zero-mean, Gaussian ensemble respecting the boundary conditions; since $\overline{T'}$ is induced by fluid motions, the natural choice is $\overline{T'}(z, 0) = 0$.}

Of the two motion-induced mean quantities, $\overline{T'}$ and $\overline{p'}$, only 
$\overline{T'}$ enters the equations for the fluctuations. Unless $\overline{p'}$ is required---it is not for our purposes---we can henceforth drop \Eq{eq:ns-av}. 
\Eqss{eq:ent-av}{z-bc-fl} and lateral boundary conditions, provide a complete description of the dynamics of $\left(\overline{T'}, \vv, p, \theta\right)$. Two general characteristics of this physical system are:

\noindent
1. The conservation form of \Eq{eq:ent-av} allows for the identification of various motion-induced---or convective---heat fluxes:\footnote{We note that, to ensure dimensional consistency, \Eqss{eq:ent-av}{eq:f-conv} should be multiplied by the constant quantity $\rho c_p$ (where $c_p$ is the specific heat at constant pressure) to obtain an equation for the entropy density. In this case, the enthalpy flux is $\rho c_p\overline{\theta v_z}$, with similar expressions for the other convective heat fluxes. However, in this paper, we will use the terminology introduced in \Eqss{eq:f-en}{eq:f-conv}.}
\begin{subequations}
\begin{align} 
F_{\rm en}(z, t) &\;=\; \overline{\theta v_z}\,,
\qquad \mbox{Enthalpy Flux}\,;
\label{eq:f-en}\\[1ex]
F'(z, t) &\;=\; -\chi\frac{\partial\overline{T'}}{\partial z}\,,
\qquad \mbox{Motion-induced Conductive Flux}\,;
\label{eq:f-p}\\[1ex]
F_{\rm c}(z, t) &\;=\; F_{\rm en} \;+\; F'\,,\qquad \mbox{Total Convective Flux}\,.
\label{eq:f-conv}
\end{align}
\end{subequations}

\noindent
2. Inheriting the reflection symmetry about the mid-plane expressed by \Eq{eq:dyn-symm}, this system is invariant under the combined transformation:
\beq
\{x,y,z'\} \to \{x,y, -z'\}\,;\;
\{\overline{T'}, \theta\} \to -\{\overline{T'}, \theta\}\,;\;
\{v_x,v_y,v_z, p\} \to \{v_x,v_y, -v_z, p\}\,.
\label{eq:dyn-symm2}
\eeq
The vertical profiles of horizontal averages of physical quantities---explored below in \Sec{sec:num2}---appear to broadly respect this symmetry, at least in a statistical sense at the level of single-point, first and second order moments.

\subsection{Numerical Exploration of Saturated Vertical Profiles}\label{sec:num2}

\begin{figure}[hbt!]
\begin{center}
\includegraphics[width=\columnwidth]{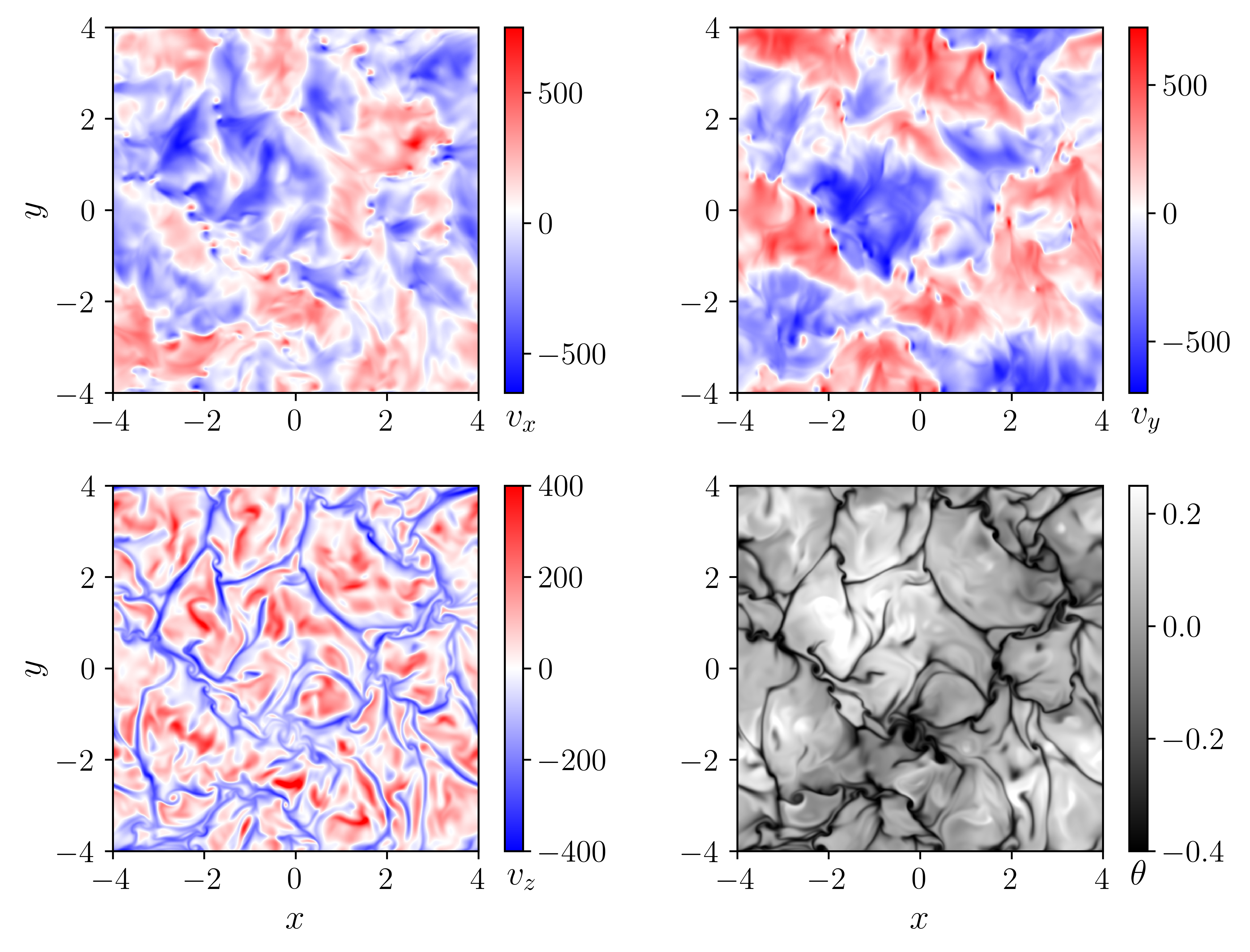}
\end{center}
\caption[]{Snapshot of velocity and temperature variations near the top boundary 
$(z=0.9)$. }
\label{flow_theta_zo9}
\end{figure}

A snapshot from the numerical simulation of \Sec{sec:num1}, taken near the top boundary 
$(z=0.9)$, is displayed in \Fig{flow_theta_zo9}. A global sense of the vertical variation of velocity and temperature fluctuations is obtained by comparision with \Fig{flow_theta_midz}, taken in the mid-plane $(z=0.5)$ e.g., in the former figure the  ``granulation'' is more sharply defined for $v_z$ and $\theta$.

\begin{figure}[hbt!]
\begin{center}
\includegraphics[width=\columnwidth]{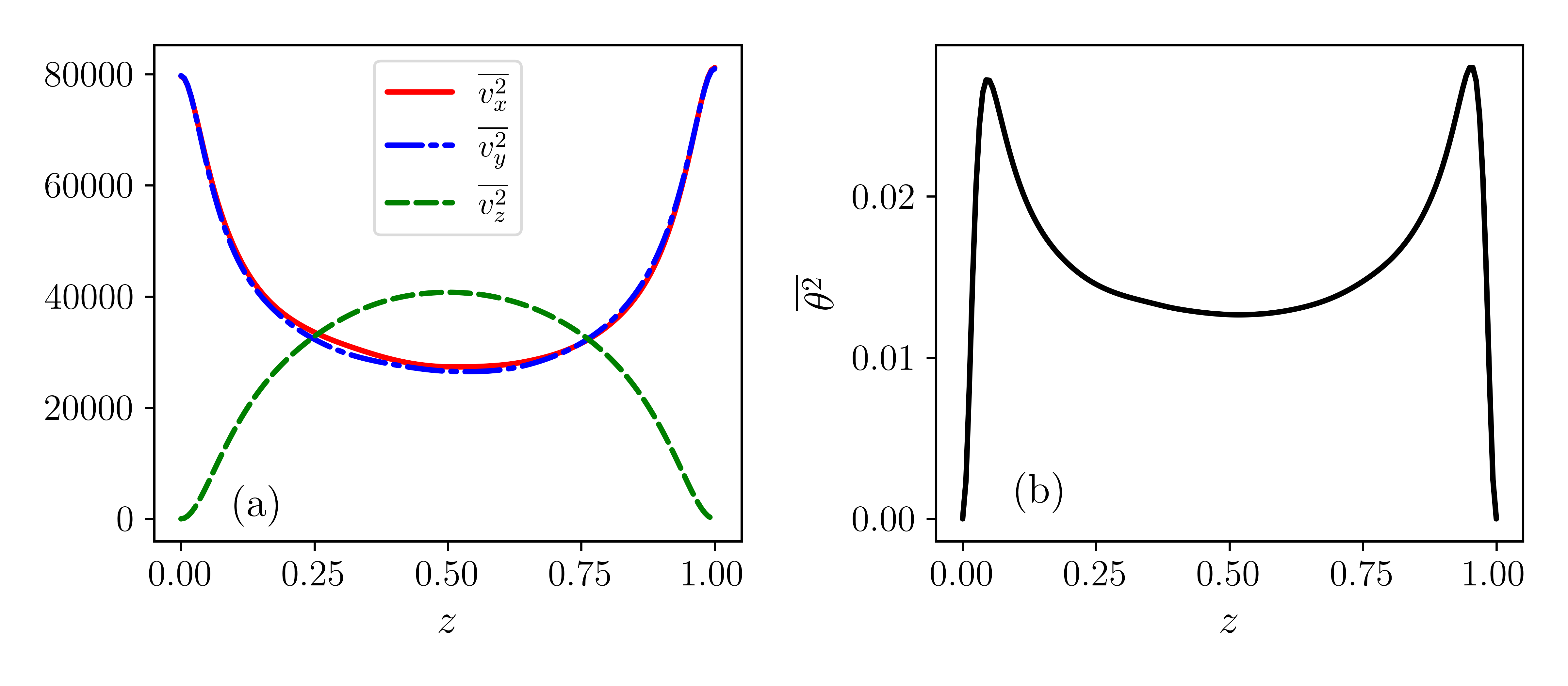}
\end{center}
\caption[]{(a) Vertical profiles of the velocity variances $\overline{v_x^2}$, 
$\overline{v_y^2}$, and  $\overline{v_z^2}$.
(b) Vertical profile of the temperature variance $\overline{\theta^2}$.
}\label{variances}
\end{figure}

When the turbulence saturates at late times $t\to\infty$, the vertical profiles of
\Eq{quad-av-non} tend to time-independent forms:
\beq
\overline{T'}(z)\,;\quad
\overline{v_x^2}(z)\,=\, \overline{v_y^2}(z)\,;\quad
\overline{v_z^2}(z)\,;\quad 
\overline{\theta^2}(z)\,;\quad
\overline{\theta v_z}(z)\,.
\label{av-non-sat}
\eeq
We consider an ensemble of velocity and temperature fields such that ensemble averages are equal to horizontal averaging. If the saturated state of the ensemble respects the 
reflection symmetry about the mid-plane of \Eq{eq:dyn-symm2}, then we expect 
$\overline{v_x^2}(z) = \overline{v_y^2}(z), \,\overline{v_z^2}(z), \,\overline{\theta^2}(z)$, and $\,\overline{\theta v_z}(z)$ to be symmetric about the mid-plane $(z=0.5)$, and $\overline{T'}(z)$ to be antisymmetric about the mid-plane. This expectation is broadly satisfied by the profiles in \Figs{variances}{fluxes}. As expected---see \Sec{sec:sym}---$\overline{v_x^2}$ and $\overline{v_y^2}$ are very nearly equal to each other throughout the $z$-interval. Averages of physical quantities over the entire fluid, or \emph{box-averages}, are equivalent to averaging over the vertical profiles. From \Fig{variances}, the four box-averaged variances are:
\beq
\left[v_x^2\right] \,\simeq\, \left[v_y^2\right] \,\simeq\, 3.9\times 10^4,\qquad
\left[v_z^2\right]  \,\simeq\, 2.8\times 10^4,\qquad
\left[\theta^2\right] \,\simeq\, 1.6\times 10^{-2}.
\label{eq:box-var}
\eeq

\begin{figure}[hbt!]
\begin{center}
\includegraphics[width=\columnwidth]{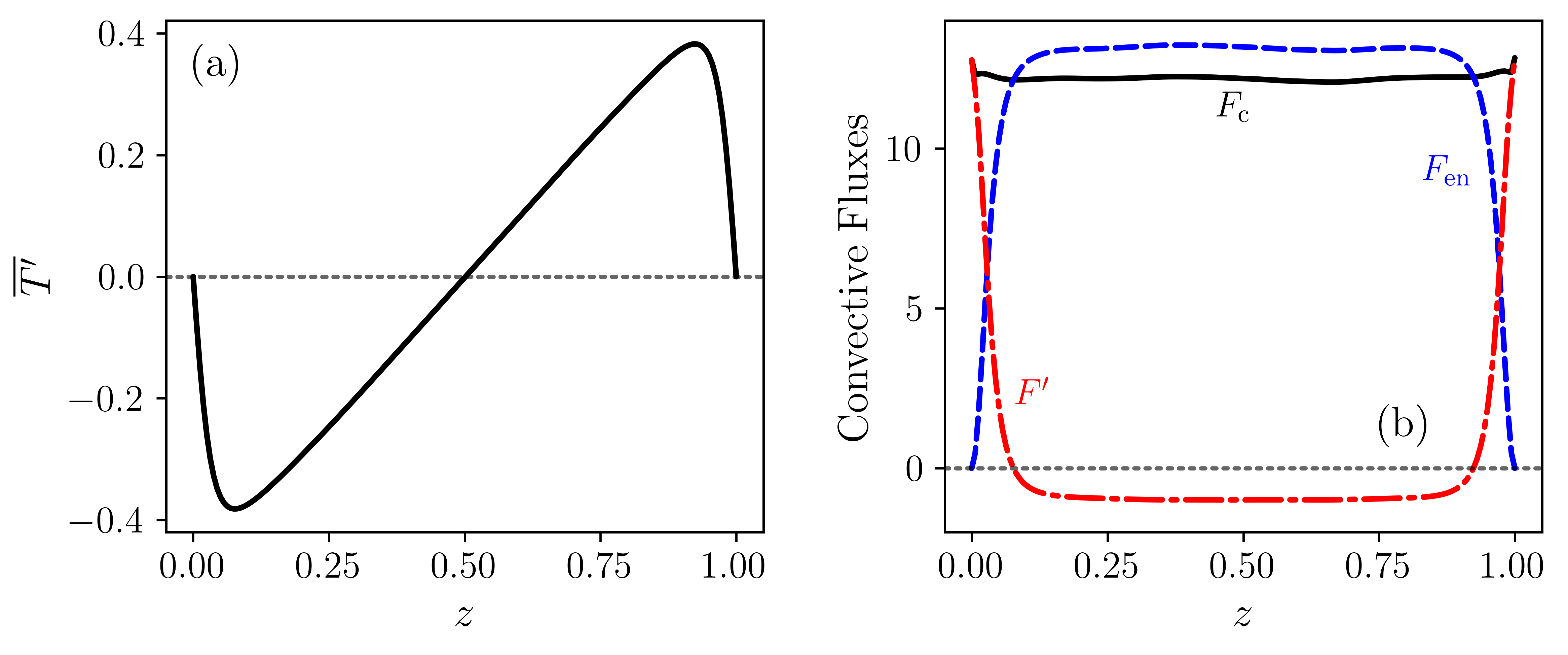}
\end{center}
\caption[]{(a) Vertical profile of $\overline{T'}$.
(b) Vertical profiles of the Convective Fluxes $F_{\rm en}, \; F',\; F_{\rm c}$.
}\label{fluxes}
\end{figure}

Since the saturated $\overline{T'}$ is time-independent, \Eq{eq:ent-av} implies that the 
total convective flux is constant. From \Eqss{eq:f-en}{eq:f-conv}, we have:
\beq
F_{\rm c} \;=\;  F_{\rm en}(z) \,+\, F'(z) \;=\; \overline{\theta v_z} \,-\, \chi\frac{\partial\overline{T'}}{\partial z} \;=\; \mbox{constant}\,.
\label{eq:f-conv2}
\eeq
\noindent
In Panel~(b) of \Fig{fluxes}, $F_{\rm en}(z)$ and $F'(z)$ vary sharply within the top and bottom boundary layers, but are nearly constant in the bulk of the fluid; however, their sum $F_{\rm c}$ is indeed nearly constant across the entire fluid. For an ensemble respecting the reflection symmetry expressed by \Eq{eq:dyn-symm2}, both $F_{\rm en}(z)$ and $F'(z)$ are expected to be symmetric about the mid-plane, which is also seen to be broadly obeyed.

The box-averaged temperature gradient is zero, because $\overline{T'} = 0$ at the top and bottom boundaries. Hence $\left[F'\right](t) = 0$ even in the fully time-independent, unsaturated phase of turbulence. Therefore, the box-averaged total convective flux is entirely due to the box-averaged enthalpy flux: $\,\left[F_{\rm c}\right]\!(t) = \left[F_{\rm en}\right]\!(t)\,$. Averaging over the saturated profiles of the convective fluxes in Panel~(b) of \Fig{fluxes}, we find: 
\beq
\left[F_{\rm en}\right] \,\simeq\, 12.1,\qquad
\left[F'\right] \,\simeq\, 0.1,\qquad
\left[F_{\rm c}\right] \,=\, \left[F_{\rm en}\right] \,+\, \left[F'\right] \,\simeq\, 
12.2\,, 
\label{eq:box-f}
\eeq
\noindent
When compared with the dimensionless reference scale for temperature-velocity correlations  $\theta_{\rm rms}v_{\rm rms}/3 \simeq 14$, we see that $\left[F_{\rm en}\right]$ is of order unity; this is because of the strong correlations between $\theta$ and $v_z$, seen in \Figs{flow_theta_midz}{flow_theta_zo9}. $\left[F'\right]$ is small, consistent with zero to within a $1\%$ margin.

\section{Fourier Expansion and Linear Modes}\label{sec:fou-lin}
Having obtained some idea of the saturated state of turbulent convection, we return to  \Eqss{eq:ent-av}{z-bc-fl}; these equations, together with lateral boundary conditions, provide a complete description of the dynamics of $\left(\overline{T'}, \vv, p, \theta\right)$. Our goal is to cast this system into a uniform modal form by choosing suitable variables; this is achieved in \Sec{sec:mod-eqns}. Here, we take the first steps by using Fourier expansions to satisfy the boundary conditions, and then we briefly review the linear modes of the theory.

\subsection{Fourier Expansion}\label{sec:fou-exp}
The vertical boundary conditions in \Eq{z-bc-fl} are satisfied by a Fourier series expansion in $z$. Using the notation $\xx_{\!\perp} = (x, y)$, we write:
\begin{align}
v_x(\xx, t) &\;=\; \sum_n w_{nx}(\xx_{\!\perp}, t)\cos(q_n z)\,,
\qquad
p(\xx, t) \;=\; \sum_n p_n(\xx_{\!\perp}, t)\cos(q_n z)\,,
\nonumber\\
v_y(\xx, t) &\;=\; \sum_n w_{ny}(\xx_{\!\perp}, t)\cos(q_n z)\,,
\qquad 
\overline{T'}(z, t) \;=\; \sum_n \tau_n(t)\sin(q_n z)\,,
\label{eq:fou-z}\\
v_z(\xx, t) &\;=\; \ii\!\sum_n w_{nz}(\xx_{\!\perp}, t)\sin(q_n z)\,,
\qquad
\theta(\xx, t) \;=\; \ii\!\sum_n \theta_n(\xx_{\!\perp}, t)\sin(q_n z)\,.
\nonumber
\end{align}
\noindent
Here $\sum_n = \sum_{n=-\infty}^{\infty}$ and $q_n = n(\pi/d)$ is the vertical wavenumber.
The $\tau_n$ are functions of $t$ only, whereas $\left(\ww_n, p_n, \theta_n\right)$ 
are functions of $(\xx_{\!\perp}, t)$. The Fourier amplitudes satisfy the following 
properties:
\begin{align}
w_{nx} &= w_{-nx} = \mbox{real}\,,\quad w_{ny} = w_{-ny} = \mbox{real}\,,\quad p_{n} = p_{-n} = \mbox{real}\,,
\nonumber\\
w_{nz} &= -w_{-nz} = \mbox{imaginary}\,,\quad \theta_{n} = -\theta_{-n} = \mbox{imaginary}\,,\quad \tau_{n} = -\tau_{-n} = \mbox{real}\,.
\label{eq:fou-amp-prop}\\
w_{0z} &= 0\,,\quad \tau_0 = 0\,,\quad \theta_0 = 0\,,\quad
\mbox{since $\left(w_{0z}, \tau_0, \theta_0\right)$ premultiply $\sin(q_0z) = 0$.}
\nonumber
\end{align}
\noindent 
The lateral boundary conditions are that either the quantities $\left(\ww_n, p_n, \theta_n\right)$ vanish as $\left|\xx_{\!\perp}\right| \to \infty$, or they are periodic functions of $\xx_{\!\perp}$. 

Toward the development of a modal theory of turbulent convection, it proves more useful to work with their horizontal Fourier transforms. Thus, we define the mode functions:
\beq
\begin{bmatrix}
\widetilde{\ww}_n\!\left(\bkap, t\right) \\
\widetilde{p}_n\!\left(\bkap, t\right) \\
\widetilde{\theta}_n\!\left(\bkap, t\right)
\end{bmatrix}
\;=\; \int 
\begin{bmatrix}\ww_n\!\left(\xx_{\!\perp}, t\right) \\
p_n\!\left(\xx_{\!\perp}, t\right) \\
\theta_n\!\left(\xx_{\!\perp}, t\right)
\end{bmatrix}
\exp\!\left(-\ii\bkap\cendot\xx_{\!\perp}\right)\dd\xx_{\!\perp}\,, 
\label{eq:mode-func-def}
\eeq
\noindent
where $\bkap = \left(\kappa_x, \kappa_y\right)$ is the horizontal wavevector. Symmetries of the mode functions follow from \Eq{eq:fou-amp-prop}:
\begin{align}
\widetilde{w}_{nx}\!\left(\bkap, t\right) &\;=\; \widetilde{w}^*_{nx}\!\left(-\bkap, t\right) \;=\; \widetilde{w}_{-nx}\!\left(\bkap, t\right)\,,
\quad\mbox{and similarly for $\widetilde{w}_{ny}$ and $\widetilde{p}_n$.}
\nonumber\\
\widetilde{w}_{nz}\!\left(\bkap, t\right) &\;=\; -\widetilde{w}^*_{nz}\!\left(-\bkap, t\right) \;=\; -\widetilde{w}_{-nz}\!\left(\bkap, t\right)\,,
\quad\mbox{and similarly for 
$\widetilde{\theta}_n$.}
\label{eq:fou-mode-prop}\\
\widetilde{w}_{0z}\!\left(\bkap, t\right) &\,=\, 0\,,\quad \widetilde{\theta}_0\!\left(\bkap, t\right) \,=\, 0\,.
\nonumber
\end{align}
\noindent
Fourier transforms are natural for the case of infinite lateral boundary conditions. For periodic lateral boundary conditions, the Fourier transforms must be specialized to Fourier series in $\xx_{\!\perp}$.

\subsection{Linear Modes}\label{sec:lin-mode}
We consider small fluctuations about hydrostatic equilibrium, neglecting the products of fluctuating quantities in the governing equations. Thus, \Eq{eq:ent-av} determining the mean temperature reduces to the source-free diffusion equation, $\partial\overline{T'}/\partial t = \chi\nabla^2 \,\overline{T'}$. With the initial condition $\overline{T'}(z, 0) = 0$, the solution is $\overline{T'}(z, t) = 0$. Consequently, the linearized \Eqss{eq:incom-fl}{eq:pres-fl} are:
\begin{subequations}
\begin{align}
\bnabla\cendot\,\vv &\;=\; 0\,,\qquad
\frac{\partial \vv}{\partial t} \;=\; -\frac{\bnabla p}{\rho} 
\,+\, g\alpha\theta\hat{z} \,+\, \nu\bnabla^2 \vv\,, 
\label{eq:lin1}\\
\frac{\partial \theta}{\partial t}  &\;=\; \beta v_z \,+\, \chi\nabla^2 \theta\,,
\qquad \frac{\nabla^2 p}{\rho} \;=\; g\alpha\frac{\partial\theta}{\partial z}\,.
\label{eq:lin2}
\end{align}
\end{subequations}
\noindent
Linearized equations for mode functions are obtained by substituting the Fourier expansions of \Eq{eq:fou-z} into \Eqs{eq:lin1}{eq:lin2} and taking horizontal Fourier transforms:
\begin{subequations}
\begin{align}
\kk_n\cendot\,\widetilde{\ww}_n &\;=\; 0\,,\qquad
\frac{\partial \widetilde{\ww}_n}{\partial t} \;=\; -\ii\kk_n\frac{\,\widetilde{p}_n\,}{\rho} \,+\, g\alpha\widetilde{\theta}_n\hat{z} \,-\, \nu k_n^2\widetilde{\ww}_n\,, 
\label{eq:lin-mod1}\\
\frac{\partial \widetilde{\theta}_n}{\partial t}  &\;=\; \beta\widetilde{w}_{nz} \,-\, \chi k_n^2 \widetilde{\theta}_n\,,
\qquad -k_n^2\frac{\widetilde{p}_n}{\rho} \;=\; \ii q_n g\alpha\widetilde{\theta}_n\,.
\label{eq:lin-mod2}
\end{align}
\end{subequations}
\noindent
Here, $\kk_n = \left(\bkap, q_n\right)$ is the three dimensional wavevector, and $k_n^2 = \kk_n\cendot\,\kk_n = \kappa^2 + q_n^2$. These linear equations admit modal solutions of the form $(\widetilde{\ww}_n, \widetilde{p}_n, \widetilde{\theta}_n) \propto \exp(\gamma_n t)$, where $\gamma_n(\bkap)$ is the growth rate. There are two kinds of linear modes: \emph{gravity modes} for which $\gamma_n(\bkap) \neq -\nu k_n^2$, and \emph{horizontal modes} for which $\gamma_n(\bkap) = -\nu k_n^2$. Since the properties of these linear modes are well-understood \citep{lss61} we summarize the essential results below.

\subsubsection{Gravity Modes: $\gamma_n(\bkap) \neq -\nu k_n^2$}\label{sec:gmode}
These are linear modes for which buoyancy acts as the restoring force, and the quantities
$\widetilde{\ww}_n$, $\widetilde{p}_n$, and $\widetilde{\theta}_n$ are generally non-zero. Furthermore, $\widetilde{\ww}_n$ and $\widetilde{\theta}_n$ are related by:
\beq
\widetilde{\ww}_n(\bkap, t) \;=\; \ea_n(\bkap)\frac{\kappa/k_n}{(\gamma_n + \nu k_n^2)}\,g\alpha\,\widetilde{\theta}_n(\bkap, t)\,,
\label{gmode-rel}
\eeq
where 
\beq
\ea_n(\bkap) \;=\; \left(-\frac{\,\kappa_x q_n\,}{\kappa\, k_n},\, -\frac{\,\kappa_y q_n\,}{\kappa\, k_n},\, \frac{\,\kappa\,}{k_n}\right)
\label{ea-def}
\eeq 
\noindent
is a unit vector which lies in the plane spanned by $\kk_n$ and $\hat{z}$ and is perpendicular to $\kk_n$. Since fluid velocities point along $\ea_n(\bkap)$, we may regard this unit vector as the polarization vector for a gravity mode with a given $\kk_n$. 
It satisfies:
\begin{align}
\ea_{nx}(\bkap) &\;=\; -\ea_{nx}(-\bkap) \;=\; -\ea_{-nx}(\bkap)\,,
\quad\mbox{and similarly for $\ea_{ny}$.}
\nonumber\\
\ea_{nz}(\bkap) &\;=\; \ea_{nz}(-\bkap) \;=\; \ea_{-nz}(\bkap)\,.
\label{eq:ea}
\end{align}

When $n=0$, we have $k_0 = \kappa$ and $\ea_0 = \hat{z}$. Then, \Eq{gmode-rel} implies that $\widetilde{w}_{0x} = \widetilde{w}_{0y} = 0$, and $\widetilde{w}_{0z} = 
\widetilde{\theta}_0/(\gamma_0 + \nu \kappa^2)$. However, from \Eq{eq:fou-mode-prop} we have  $\widetilde{w}_{0z}=0$ and $\widetilde{\theta}_0=0$. \emph{Therefore, there are no gravity modes with $n=0$}.

The growth rate satisfies a quadratic equation, whose two solutions are:
\beq
\gamma_n^{\pm}(\kappa) \;=\; -\frac{(\chi + \nu)}{2}k_n^2 \;\pm\;
\sqrt{g\alpha\beta\frac{\kappa^2}{k_n^2} \,+\, \frac{(\chi - \nu)^2}{4}k_n^4}\;\;.
\label{gamma-pm}
\eeq
\noindent
When $\beta$ is positive, both $\gamma_n^{\pm}$ are real. $\gamma_n^{-}$ is always negative, corresponding to \emph{decaying} gravity modes, and we always have
$\gamma_n^{+} \geq \gamma_n^{-}$. However, \emph{growing} gravity modes require 
$\,\gamma_n^{+}(\kappa)>0$, which occurs when the Rayleigh number exceeds a critical value 
$\ra_{\rm cr}$. This, along with the critical modes excited, is discussed in \App{app:ra-crit}. For infinite lateral boundary conditions, as well as the numerical simulation discussed in \Sec{sec:num1}, $\ra_{\rm cr} = 657.51\ldots\,$ and the critical modes have $n=\pm 1$ and $\kappa d = \pi/\sqrt{2}$. While the critical modes are isotropic in the horizontal directions for the infinite case, they have $\kappa_x = \kappa_y = \pm 4\pi/L$ for the numerical simulation. Since the numerical simulation had $\ra = 10^6$, which is much larger than $\ra_{\rm cr}$, we expect many growing modes to be excited in addition to the critical mode.

\subsubsection{Horizontal Modes: $\gamma_n(\bkap) = -\nu k_n^2$}\label{sec:hmode}
These are linear modes for which buoyancy is not a restoring force; the quantities
$\widetilde{w}_{nz}$, $\widetilde{p}_n$, and $\widetilde{\theta}_n$ vanish. In the absence of buoyancy and temperature fluctuations, the modes decay viscously. Fluid elements move horizontally, with $\widetilde{\ww}_n(\bkap, t) \propto \eu(\bkap)$, where 
\beq
\eu(\bkap) \;=\; \mathbf{\hat{\kk}}_n\,\cross\,\ea_n(\bkap) \;=\; \left(\frac{\,\kappa_y\,}{\kappa},\, -\frac{\,\kappa_x\,}{\kappa},\, 0\right)
\label{eu-def}
\eeq
\noindent
is a unit vector perpendicular to both $\kk_n$ and $\ea_n(\bkap)$, and it is independent of $n$. Since fluid velocities point along $\pm\eu(\bkap)$, we may regard this unit vector as the polarization vector for a horizontal mode with a given $\bkap$. It satisfies:
\beq
\eu(\bkap) \;=\; -\eu(-\bkap)\,.
\label{eq:eu}
\eeq
\noindent
Although the polarization vectors are independent of $n$, horizontal modes with different $n$ decay at different rates, given by $\gamma^u_n(\kappa) = -\nu k_n^2$. In contrast with gravity modes, all integer values of $n$ are allowed; notably, the $n=0$ horizontal mode exhibits $z$--independent velocities.

\section{Nonlinear Development}\label{sec:non-dev}
Having taken care of the vertical boundary conditions in \Eq{z-bc-fl} through Fourier expansions and reviewed linear theory, we are in a position to deal with \Eqss{eq:ent-av}{z-bc-fl} and lateral boundary conditions in generality.

\subsection{The Mean Temperature}\label{sec:fou-amp}
We solve \Eq{eq:ent-av} for $\overline{T'}$ by using the Fourier expansions of \Eq{eq:fou-z}. The enthalpy flux is:
\beq
\overline{\theta v_z}(z, t) \;=\; \sum_n \Lambda_n(t)\cos(q_n z)\,,\quad
\mbox{where}\quad \Lambda_n(t) \;=\; \sum_{\ell, m}\delta_{\ell m, n}\,\overline{\theta_\ell w_{mz}}(t)\,.
\label{eq:f-en-fou}
\eeq
\noindent
Here $\sum_{\ell,m} = \sum_{\ell = -\infty}^{\infty}\sum_{m = -\infty}^{\infty}\,$, and 
$\delta_{\ell m, n}$ is the Kronecker delta function, equal to $1$ when $\ell + m = n$ and $0$ otherwise. Then, $\tau_n(t)$ obeys the following ordinary differential equation:
\beq
\frac{\dd\tau_n}{\dd t} \;+\; \chi q_n^2\tau_n \;=\; 
q_n \Lambda_n(t)\,.
\label{eq:taun}
\eeq
\noindent
With the initial condition $\tau_n(0)=0$, the solution is:
\beq
\tau_n(t) \;=\; 
\begin{cases} 
   \; 0 & \quad \text{if}\quad n = 0,\\
   \; \displaystyle q_n\int_0^t\exp\!{\left[-\chi q_n^2(t-t')\right]}\,
   \Lambda_n(t')\,\dd t' & \quad \text{otherwise.}
\end{cases}
\label{eq:taun-soln}
\eeq
\noindent
Hence, the mean temperature has been expressed in terms of the fluctuating Fourier amplitudes. 

At late times, $t\to\infty$, averages like $\overline{\theta_\ell w_{mz}}(t')$ tend to constant values, $\overline{\theta_\ell w_{mz}}^{(\rm s)}$. Then, $\tau_n(t)$ in 
\Eq{eq:taun-soln} saturates to:
\beq
\tau_n^{(\rm s)} \;=\; 
\begin{cases} 
   \; 0 & \quad \text{if}\quad n = 0,\\
   \; \displaystyle \frac{\Lambda_n^{(\rm s)}}{\chi q_n}
   & \quad \text{otherwise.}
\end{cases}
\label{eq:taun-sat}
\eeq
\noindent
This saturated form will be used in \Sec{sec:self} to simplify the nonlinear problem, from being non-local in time---because of the time-integral in \Eq{eq:taun-soln}---to a local form depending parametrically on $\Lambda_n^{(\rm s)}$.

\subsection{Nonlinear Equations for Fluctuations}\label{sec:mod-func}
Nonlinear equations for the fluctuation Fourier amplitudes, $\ww_n(\xx_{\!\perp}, t)$ and 
$\theta_n(\xx_{\!\perp}, t)$, are derived by substituting the Fourier expansions of \Eq{eq:fou-z} into \Eqss{eq:incom-fl}{eq:ent-fl}, and equating the coefficients of 
$\cos(q_n z)$ and $\sin(q_n z)$, respectively, on both sides of the equations. Using the notation $\ww_{n\perp} = \left(w_{nx}, w_{ny}\right)$ and 
$\bnabla_{\!\!\perp} = \left(\partial_x, \partial_y\right)\,$, we have:
\begin{subequations}
\begin{align}
&\bnabla_{\!\!\perp}\cendot\,\ww_{n\perp} \;+\; \ii q_n w_{nz} \;=\; 0\,,
\label{eq:wn-incom}\\[1ex]
\frac{\partial \ww_{n\perp}}{\partial t} \;+\; \sum_{\ell, m}\delta_{\ell m, n}
&\left[\,\partial_x\!\left(w_{m x}\ww_{\ell\perp}\right) \,+\, \partial_y\!\left(w_{m y}\ww_{\ell\perp}\right) \,+\, \ii q_n\!\left(w_{m z}\ww_{\ell\perp} - \overline{w_{m z}\ww_{\ell\perp}}\right)\right] 
\nonumber\\
&\;=\; -\frac{\bnabla_{\!\!\perp} p_n}{\rho} \,+\, \nu\!\left[\nabla_{\!\!\perp}^2
\ww_{n\perp} - q_n^2 \ww_{n\perp}\right],
\label{eq:wnperp}\\[1ex]
\frac{\partial w_{nz}}{\partial t} \;+\; \sum_{\ell, m}\delta_{\ell m, n}
&\left[\,\partial_x\!\left(w_{m x}w_{\ell z}\right) \,+\, \partial_y\!\left(w_{m y}w_{\ell z}\right) \,+\, \ii q_n\!\left(w_{m z}w_{\ell z} - \overline{w_{m z}w_{\ell z}}\right) 
\right] 
\nonumber\\
&\;=\; -\ii q_n\frac{p_n}{\rho} \,+\, g\alpha\theta_n \,+\,
\nu\!\left[\nabla_{\!\!\perp}^2 w_{nz} - q_n^2 w_{nz}\right],
\label{eq:wnz}\\[1ex]
\frac{\partial \theta_n}{\partial t} \;+\; \sum_{\ell, m}\delta_{\ell m, n}
&\left[\,\partial_x\!\left(w_{m x}\theta_\ell\right) \,+\, \partial_y\!\left(w_{m y}\theta_\ell\right) \,+\, \ii q_n\!\left(w_{m z}\theta_\ell - \overline{w_{m z}\theta_\ell}\right)\right]\nonumber\\
&\;=\; \beta w_{nz} \,-\, \sum_{\ell, m}\delta_{\ell m, n}\,q_m\tau_m(t) w_{\ell z}
\,+\, \chi\!\left[\nabla_{\!\!\perp}^2 \theta_n - q_n^2 \theta_n\right].
\label{eq:tn}
\end{align}
\end{subequations}
\noindent
In deriving the nonlinear terms involving $\delta_{\ell m, n}$, we have used the properties listed in \Eq{eq:fou-amp-prop}.\footnote{We note that the $\overline{w_{m z}\ww_{\ell\perp}}$ terms in \Eq{eq:wnperp} are superfluous, because isotropy in the horizontal directions---see \Eq{quad-av-zero}---gives $\overline{v_z v_x} = \overline{v_z v_y} = 0$, which implies $\sum_{\ell, m}\delta_{\ell m, n}\overline{w_{m z}\ww_{\ell\perp}} = \mathbf{0}\,.$ Inclusion of these terms facilitates the writing of all components of \Eq{eq:wn-mod} in a uniform manner.}

By taking horizontal Fourier transforms of \Eqss{eq:wn-incom}{eq:tn}, we obtain nonlinear equations for the mode functions $\widetilde{\ww}_n(\bkap, t)$ and $\widetilde{\theta}_n(\bkap, t)$. Henceforth, we will deal with functions of $\bkap$, rather than 
$\xx_{\!\perp}$. It proves convenient to form an \emph{ensemble of modes} such that the horizontal averages of functions of $\xx_{\!\perp}$ translate to ensemble averages, denoted by $\bra{\cdots}$, of the corresponding horizontal Fourier transforms. 
\begin{itemize}
\item Thus, $\bra{\widetilde{\ww}_n(\bkap, t)} = \mathbf{0}$, $\,\bra{\widetilde{p}_n(\bkap, t)} = 0$, and $\,\bra{\widetilde{\theta}_n(\bkap, t)} = 0$. 

\item If $f(\xx_{\!\perp}, t)$ and $g(\xx_{\!\perp}, t)$ are two fluctuating quantities, statistical homogeneity in the horizontal directions implies that $\overline{f(\xx_{\!\perp}, t)g(\xx'_\perp, t)}$ is a function only of $(\xx_{\!\perp} - \xx'_\perp)$. This translates to $\bra{\widetilde{f}(\bkap, t) \widetilde{g}^*(\bkap', t)} = P(\bkap)\delta(\bkap - \bkap')$, where $\,\delta\,$ is the Dirac delta function and $P$ is the cross-power spectrum. 
\end{itemize}

Using the shorthand notation $\widetilde{\ww}_{n\kappa} = \widetilde{\ww}_n(\bkap, t)$, $\widetilde{\ww}_{n1} = \widetilde{\ww}_n(\bkap_1, t)$, and 
$\widetilde{\ww}_{n2} = \widetilde{\ww}_n(\bkap_2, t)$---and similar notation for the 
$\widetilde{\theta}_n(\bkap, t)$ and $\widetilde{p}_n(\bkap, t)$ terms---we have:
\begin{subequations}
\begin{align}
&\kk_n\cendot\,\widetilde{\ww}_{n\kappa} \;=\; 0\,,
\label{eq:wn-in-mod}\\[1em]
\frac{\partial \widetilde{\ww}_{n\kappa}}{\partial t} \,+\, \ii\sum_{\ell, m}&\delta_{\ell m, n}\!\!\int\!\left[\,\widetilde{\ww}_{\ell 1}\left(\kk_n\cendot\,\widetilde{\ww}_{m 2}\right) \,-\, \bbra{\widetilde{\ww}_{\ell 1}\left(\kk_n\cendot\,\widetilde{\ww}_{m 2}\right)}\,\right]\delta(12,\bkap)
\frac{\dd\bkap_1\,\dd\bkap_2}{(2\pi)^2}
\nonumber\\
&\;=\; -\ii\kk_n\frac{\widetilde{p}_{n\kappa}}{\rho} \;+\; g\alpha\widetilde{\theta}_{n\kappa}\hat{z} \;-\; \nu k_n^2\,\widetilde{\ww}_{n\kappa}\,,
\label{eq:wn-mod}\\[1em]
\frac{\partial \widetilde{\theta}_{n\kappa}}{\partial t} \,+\, \ii\sum_{\ell, m}&\delta_{\ell m, n}\!\!\int
[\,\widetilde{\theta}_{\ell 1}\left(\kk_n\cendot\,\widetilde{\ww}_{m 2}\right) \,-\, \bra{\widetilde{\theta}_{\ell 1}\left(\kk_n\cendot\,\widetilde{\ww}_{m 2}\right)}\,]\delta(12,\bkap)
\frac{\dd\bkap_1\,\dd\bkap_2}{(2\pi)^2}
\nonumber\\
&\;=\; \beta\!\left(\hat{z}\cendot\widetilde{\ww}_{n\kappa}\right) \;-\; 
\sum_{\ell, m}\delta_{\ell m, n}\,q_m\tau_m(t)\!\left(\hat{z}\cendot\widetilde{\ww}_{\ell\kappa}\right) \;-\; \chi k_n^2\,\widetilde{\theta}_{n\kappa}\,.
\label{eq:tn-mod}
\end{align}
\end{subequations}
\noindent
Here, and henceforth, we use the notation $\delta(12,\bkap) = \delta(\bkap_1 + \bkap_2 - \bkap)$. 

The purpose of $p_n(\xx_{\!\perp}, t)$, which occurs in \Eqs{eq:wnperp}{eq:wnz}, and its Fourier transform $\widetilde{p}_n(\bkap, t)$ in \Eq{eq:wn-mod}, is to maintain the incompressibility conditions in \Eqs{eq:wn-incom}{eq:wn-in-mod}, respectively. It is not necessary to derive the equations satisfied by these quantities because the incompressibility condition can be satisfied directly by the transverse velocity decomposition, as explained below in \Sec{sec:decomp}.

The passage from \Eqss{eq:wn-incom}{eq:tn} to \Eqss{eq:wn-in-mod}{eq:tn-mod} is an improvement in form. In the former equations, the spatial derivatives appear in a non-uniform manner as $\{\partial_x, \partial_y, \ii q_n\}$, necessitating separate equations for $\ww_{n\perp}$ and $w_{nz}$. In the latter equations, spatial derivatives 
are treated uniformly as $\ii\kk_n \equiv \ii\{\kappa_x, \kappa_y, q_n\}$, and a single equation for $\widetilde{\ww}_{n\kappa}$ suffices. However, some essential non-uniformity remains in \Eqss{eq:wn-in-mod}{eq:tn-mod}, in that the mode functions,  
$\{\widetilde{\ww}_{n\kappa}, \widetilde{\theta}_{n\kappa}\}$, are a mixture of vector and scalar quantities. The transverse velocity decomposition of \Sec{sec:decomp} takes the first step toward overcoming this remaining non-uniformity.

\subsubsection{Self-Consistency}\label{sec:self}
In the term $\sum_{\ell, m}\delta_{\ell m, n}\,q_m\tau_m(t)\!\left(\hat{z}\cendot\widetilde{\ww}_{\ell\kappa}\right)$ on the right-hand side of \Eq{eq:tn-mod}, 
the quantity $\tau_m(t)$ is defined in \Eq{eq:taun-soln} by a time-integral over 
$\Lambda_m(t')$, where
\beq
\Lambda_m(t) \;=\; \sum_{i, j}\delta_{i j, m}\,\overline{\theta_i w_{jz}}(t)\,.
\nonumber
\eeq
The dependence of $\tau_m(t)$ on the second-order correlator $\overline{\theta_i w_{jz}}(t)$ or its Fourier transform $\bra{\widetilde{\theta}_{i}(\bkap)\widetilde{\ww}^*_{j}(\bkap')}(t)$, makes $\tau_m(t)$ second-order. Hence, the term $\sum_{\ell, m}\delta_{\ell m, n}\,q_m\tau_m(t)\!\left(\hat{z}\cendot\widetilde{\ww}_{\ell\kappa}\right)$ in \Eq{eq:tn-mod} is formally of third order in the fields $\widetilde{\theta}$ and 
$\widetilde{\ww}$. \Eqss{eq:wn-in-mod}{eq:tn-mod}, together with the definitions of $\tau_m(t)$ and $\Lambda_m(t)$, constitute a self-consistent set of equations. 

However, these equations are nonlocal in time, because $\tau_m(t)$ is given by \Eq{eq:taun-soln} in terms of a time integral over the fluctuations. We are primarily interested in the long-time limit, in which the turbulence is approaching saturation and 
$\tau_m(t)$ has nearly attained a constant value, $\tau_m^{(\rm s)}$, given by \Eq{eq:taun-sat}. Then,  
\beq
\sum_{\ell, m}\delta_{\ell m, n}\,q_m\tau_m(t)\!\left(\hat{z}\cendot\widetilde{\ww}_{\ell\kappa}\right) \;\to\;
\frac{1}{\chi}\sum_{\ell, m \neq 0}\delta_{\ell m, n}\,
\Lambda_m^{(\rm s)}
\!\left(\hat{z}\cendot\widetilde{\ww}_{\ell\kappa}\right)\,,
\label{eq:replace}
\eeq
\noindent
where $\sum_{\ell, m\neq 0}$ is sum over all $\ell$ and $m$, excluding $m=0$, because 
$\tau_0 = 0$. The $\Lambda_m^{(\rm s)}$, including $\Lambda_0^{(\rm s)}$, are saturated values of the vertical profile coefficients of the enthalpy flux, defined in \Eq{eq:f-en-fou}. We write:
\beq
\Lambda_m^{(\rm s)} \;=\; \sum_{i, j}\delta_{i j, m}\,\overline{\theta_i w_{jz}}^{(\rm s)}
\;=\; \sum_{i, j}\delta_{i j, m}\!\!\int {\cal P}^{(\rm s)}_{ij}(\kappa)\,\kappa\,\dd\kappa\,,
\label{eq:lamb-spec}
\eeq
where ${\cal P}^{(\rm s)}_{ij}(\kappa)$ are saturated cross-spectra of temperature-vertical velocity fluctuations, which are defined in terms of two-point correlators by, 
\beq
\hat{z}\cendot\bra{\widetilde{\theta}_{i}(\bkap)\widetilde{\ww}^*_{j}(\bkap')}^{(\rm s)} \;=\; (2\pi)^3\, {\cal P}^{(\rm s)}_{ij}(\kappa)\,\delta(\bkap - \bkap')\,.
\label{eq:spec}
\eeq
\Eqss{eq:wn-in-mod}{eq:tn-mod}, together with the replacements \Eqs{eq:replace}{eq:lamb-spec}, yield a self-consistent system of equations that simultaneously determines the mean temperature profile as well as the fluctuation spectra at saturation. Being local in time, the equations are valid when the convective turbulence is close to saturation. 

\subsection{Craya-Herring Decomposition}\label{sec:decomp}
The linear modes of \Sec{sec:lin-mode} consist of gravity modes and horizontal modes, 
with polarization vectors $\ea_n(\bkap)$ and $\eu(\bkap)$, respectively. 
$\{\ea_{n\kappa}, \,\eu_\kappa, \,\mathbf{\hat{\kk}}_n\}$ forms a right-handed, orthonormal set of vectors, which is identical to the Craya-Herring basis \citep{cra58,her74} enabled by the symmetry of the problem. Hence, $\{\ea_{n\kappa}, \,\eu_\kappa\}$ is a set of basis vectors in the plane transverse to $\kk_n$. Since \Eq{eq:wn-in-mod} implies that 
$\widetilde{\ww}_{n\kappa}$ is transverse to $\kk_n$, we can always decompose 
\beq
\widetilde{\ww}_n(\bkap, t) \;=\; \ii\ea_n(\bkap)a_n(\bkap, t) \;+\; 
\ii\eu(\bkap)u_n(\bkap, t)\,,
\label{eq:w-decomp}
\eeq
\noindent
where $a_n(\bkap, t)$ and $u_n(\bkap, t)$ are the velocity amplitudes of gravity modes 
and horizontal modes, respectively, in the fully nonlinear case. We also write
\beq
\widetilde{\theta}_n(\bkap, t) \;=\; \ii c_n(\bkap, t)\,,
\label{eq:t-decomp}
\eeq
\noindent
where $c_n(\bkap, t)$ is the temperature amplitude of the gravity modes. If we are given 
$\widetilde{\ww}_{n\kappa}$ and $\widetilde{\theta}_{n\kappa}$, we can recover $a_{n\kappa}$, $c_{n\kappa}$, and $u_{n\kappa}$ as follows:
\beq
a_{n\kappa} \,=\, -\ii\ea_{n\kappa}\cendot  \widetilde{\ww}_{n\kappa}\,,\qquad
c_{n\kappa} \,=\, -\ii\widetilde{\theta}_{n\kappa}\,,\qquad
u_{n\kappa} \,=\, -\ii\eu_\kappa\cendot  \widetilde{\ww}_{n\kappa}\,.
\label{eq:recov}
\eeq
\noindent
Since $\bra{\widetilde{\ww}_{n\kappa}}=0$ and $\bra{\widetilde{\theta}_{n\kappa}}=0$, we have $\bra{a_{n\kappa}} = \bra{u_{n\kappa}} = 0$, and $\bra{c_{n\kappa}} = 0$.

The symmetry properties in \Eq{eq:fou-mode-prop}, \Eqs{eq:ea}{eq:eu} imply the following: 
\begin{align}
a_n\!\left(\bkap, t\right) &\;=\; a^*_n\!\left(-\bkap, t\right) \;=\; -a_{-n}\!\left(\bkap, t\right)\,,\quad\mbox{and similarly for $c_n$.}
\nonumber\\
u_n\!\left(\bkap, t\right) &\;=\; u^*_n\!\left(-\bkap, t\right) \;=\; u_{-n}\!\left(\bkap, t\right)\,,\label{eq:acu-symm}\\
a_0\!\left(\bkap, t\right) &\,=\, 0\,,\quad 
c_0\!\left(\bkap, t\right) \,=\, 0\,.
\nonumber
\end{align}

Evolution equations for $a_{n\kappa}$, $c_{n\kappa}$, and $u_{n\kappa}$ can be derived by substituting \Eqs{eq:w-decomp}{eq:t-decomp} into \Eqss{eq:wn-in-mod}{eq:tn-mod}, and using \Eq{eq:recov}. We note that the incompressibility condition, \Eq{eq:wn-in-mod}, is automatically satisfied, as this was the purpose of the transverse decomposition in \Eq{eq:w-decomp}. A related fact is that when \Eq{eq:wn-mod} is dotted with $\ea_{n\kappa}$ and $\eu_{\kappa}$ to get equations for $a_{n\kappa}$ and $u_{n\kappa}$, respectively, the pressure term $\widetilde{p}_{n\kappa}$ drops out since $\kk_n$ is perpendicular to both $\ea_{n\kappa}$ and $\eu_{\kappa}$. Thus: 
\begin{subequations}
\begin{align}
\frac{\partial a_{n\kappa}}{\partial t} &\;=\; g\alpha\frac{\kappa}{k_n}c_{n\kappa} \;-\; \nu k_n^2 a_{n\kappa} \;+\; \Gamma^a_{n\kappa} \,-\, \bbra{\Gamma^a_{n\kappa}}\,,
\label{eq:a-evol}\\[1ex]
\frac{\partial c_{n\kappa}}{\partial t} &\;=\; \beta\frac{\kappa}{k_n}a_{n\kappa} \;-\; \chi k_n^2 c_{n\kappa} \;+\; \Gamma^c_{n\kappa} \,-\, \bbra{\Gamma^c_{n\kappa}}
\;-\; \sum_\ell \Theta_{n\ell}(\kappa)a_{\ell\kappa}\,,
\label{eq:c-evol}\\[1ex]
\frac{\partial u_{n\kappa}}{\partial t} &\;=\; -\nu k_n^2 u_{n\kappa}
\;+\; \Gamma^u_{n\kappa} \,-\, \bbra{\Gamma^u_{n\kappa}}\,,
\label{eq:u-evol}
\end{align}
\end{subequations}
\noindent
Here, $\{\Gamma^a_{n\kappa}, \Gamma^c_{n\kappa}, \Gamma^u_{n\kappa}\}$ are quadratic functionals of $\{a, c, u\}$: 
\begin{subequations}
\begin{align}
\Gamma^a_{n\kappa} &\;=\; 
\sum_{\ell, m} \,\delta_{\ell m, n}\!\!\int\!\left[ La_{\ell 1}a_{m2} \,+\, Ma_{\ell 1}u_{m2} \,+\, Uu_{\ell 1}u_{m2}\right]\delta(12,\bkap)\frac{\dd\bkap_1\dd\bkap_2}{(2\pi)^2}\,,
\label{eq:Gam-a}\\[1ex] 
\Gamma^c_{n\kappa} &\;=\; 
\sum_{\ell, m}\delta_{\ell m, n}\!\!\int\!\left[Pc_{\ell 1}a_{m2} \,+\, Qc_{\ell 1}u_{m2}\right]\delta(12,\bkap)\frac{\dd\bkap_1\dd\bkap_2}{(2\pi)^2}\,,
\label{eq:Gam-c}\\[1ex]
\Gamma^u_{n\kappa} &\;=\; 
\sum_{\ell, m} \,\delta_{\ell m, n}\!\!\int\!\left[Ra_{\ell 1}a_{m2} \,+\, Sa_{\ell 1}u_{m2} \,+\, Vu_{\ell 1}u_{m2}\right]\delta(12,\bkap)\frac{\dd\bkap_1\dd\bkap_2}{(2\pi)^2}\,.
\label{eq:Gam-u}
\end{align}
\end{subequations}
\noindent
The eight coefficients $\{L, M, \ldots , S, V\}$ are given in \Eq{eq:coeffs-acu}. 
\beq
\Theta_{n\ell}(\kappa) \;=\; 
\begin{cases} 
   \; 0 & \quad \text{if}\quad \ell = n,\\[1ex]
   \; \displaystyle \frac{\Lambda_{(n-\ell)}^{(\rm s)}}{\chi}\frac{\kappa}{k_\ell}

  & \quad \text{otherwise,}
\end{cases}
\label{eq:Theta-sat}
\eeq
are simple functions with parameters $\Lambda_{(n-\ell)}^{(\rm s)}$, that have to 
be related self-consistently to saturated spectra; see \Sec{sec:self}. 

\section{Modal Equations and Energy Pathways}\label{sec:mod-path}
\subsection{Modal Equations}\label{sec:mod-eqns}
The Craya-Herring decomposition of the previous section was a key step enabling a uniform 
description in terms of the variables $\left\{a_{n\kappa}, c_{n\kappa}, u_{n\kappa}\right\}$, avoiding an uneven mixture of scalar and vector variables. However, 
\Eqss{eq:a-evol}{eq:u-evol} governing these quantities are not yet in a form that is 
convenient for further exploration. The main reason arises from the form of 
the linear terms of these equations:
\beq 
\partial_t a_{n\kappa} = g\alpha\frac{\kappa}{k_n}c_{n\kappa} - \nu k_n^2 a_{n\kappa}\,,\quad
\partial_t c_{n\kappa} = \beta\frac{\kappa}{k_n}a_{n\kappa} - \chi k_n^2 c_{n\kappa}\,,\quad
\partial_t u_{n\kappa} = -\nu k_n^2 u_{n\kappa}\,.
\label{eq:acu-lin}
\eeq
\noindent
The $u_{n\kappa}$ are linear horizontal eigenmodes, but $a_{n\kappa}$ and  $c_{n\kappa}$ are not linear gravity eigenmodes, because they satisfy mutually coupled equations. 

We define the linear combinations of $a_{n\kappa}$ and  $c_{n\kappa}$:
\beq
\varphi^+_n(\bkap, t) \;=\; a_n(\bkap, t) \;+\; \lambda^+_{n\kappa}c_n(\bkap, t)\,,
\qquad
\varphi^-_n(\bkap, t) \;=\; a_n(\bkap, t) \;+\; \lambda^-_{n\kappa}c_n(\bkap, t)\,,
\label{eq:phi-def}
\eeq
\noindent
where the coefficients,
\beq
\lambda^+_{n\kappa} \;=\; \frac{\gamma^+_{n\kappa} + \nu k_n^2}{\beta (\kappa/k_n)}\,,
\qquad 
\lambda^-_{n\kappa} \;=\; \frac{\gamma^-_{n\kappa} + \nu k_n^2}{\beta (\kappa/k_n)}\,,\label{eq:lamb-pm}
\eeq
\noindent
are defined in terms of the gravity mode growth rates $\gamma^\pm_n(\kappa)$ of \Eq{gamma-pm}. In the limit of linear theory, \Eq{eq:acu-lin} implies that:
\beq
\partial_t \varphi^+_{n\kappa} =  \gamma^+_{n\kappa}\varphi^+_{n\kappa}\,,\qquad
\partial_t \varphi^-_{n\kappa} =  \gamma^-_{n\kappa}\varphi^-_{n\kappa}\,,\qquad
\partial_t u_{n\kappa} = -\nu k_n^2 u_{n\kappa}\,.
\label{eq:pmu-lin}
\eeq
\noindent
Therefore, $\{\varphi^\pm_{n\kappa}, u_{n\kappa}\}$ reduces to three types of linear eigenmodes---growing gravity modes, decaying gravity modes, and horizontal modes, respectively---as discussed in \Sec{sec:lin-mode}. Since the definitions of 
$\{\varphi^\pm_{n\kappa}, u_{n\kappa}\}$ are valid in the fully nonlinear case, this set of functions---referred to as \emph{mode amplitudes}---is a convenient choice for a modal description of turbulent Boussinesq convection.

Modal equations governing $\varphi^\pm_{n\kappa}$ are derived by multiplying \Eq{eq:c-evol} by $\lambda^\pm_{n\kappa}$ and adding the result to \Eq{eq:a-evol}. The task is complete when $a_{n\kappa}$ and  $c_{n\kappa}$ are expressed in terms of 
$\varphi^\pm_{n\kappa}\,$ by inverting \Eq{eq:phi-def}:
\beq
a_{n\kappa} \;=\; \frac{\lambda^+_{n\kappa}\varphi^-_{n\kappa} \,-\, \lambda^-_{n\kappa}\varphi^+_{n\kappa}}{\lambda^+_{n\kappa} - \lambda^-_{n\kappa}}\,,\qquad
c_{n\kappa} \;=\; \frac{\varphi^+_{n\kappa} - \varphi^-_{n\kappa}}{\lambda^+_{n\kappa} - \lambda^-_{n\kappa}}\,.
\label{eq:ac-inv}
\eeq
\noindent
Together with \Eq{eq:u-evol}, the equations governing the three mode amplitudes are:
\begin{subequations}
\begin{align}
\frac{\partial \varphi^+_{n\kappa}}{\partial t} &\;=\;  \gamma^+_{n\kappa}\varphi^+_{n\kappa} \;+\; \Gamma^+_{n\kappa} \,-\, \bbra{\Gamma^+_{n\kappa}}
\;-\; \lambda^+_{n\kappa}\sum_{\ell}\Theta_{n\ell}(\kappa)\,
\frac{\lambda^+_{\ell\kappa}\varphi^-_{\ell\kappa} \,-\, \lambda^-_{\ell\kappa}\varphi^+_{\ell\kappa}}{\lambda^+_{\ell\kappa} - \lambda^-_{\ell\kappa}}\,,
\label{eq:phip-evol}\\[1ex]
\frac{\partial \varphi^-_{n\kappa}}{\partial t} &\;=\;  \gamma^-_{n\kappa}\varphi^-_{n\kappa} \;+\; \Gamma^-_{n\kappa} \,-\, \bbra{\Gamma^-_{n\kappa}}
\;-\; \lambda^-_{n\kappa}\sum_{\ell}\Theta_{n\ell}(\kappa)\,\frac{\lambda^+_{\ell\kappa}\varphi^-_{\ell\kappa} \,-\, \lambda^-_{\ell\kappa}\varphi^+_{\ell\kappa}}{\lambda^+_{\ell\kappa} - \lambda^-_{\ell\kappa}}\,,
\label{eq:phim-evol}\\[1ex]
\frac{\partial u_{n\kappa}}{\partial t} &\;=\; -\nu k_n^2 u_{n\kappa}
\;+\; \Gamma^u_{n\kappa} \,-\, \bbra{\Gamma^u_{n\kappa}}\,.
\label{eq:u-evol-new}
\end{align}
\end{subequations}
\noindent
Here, the quadratic functionals, $\,\Gamma^\pm_{n\kappa} \,=\, \Gamma^a_{n\kappa} \,+\, \lambda^\pm_{n\kappa}\Gamma^c_{n\kappa}\,$, are linear combinations of 
$\Gamma^a_{n\kappa}$ and $\Gamma^c_{n\kappa}$ in \Eqs{eq:Gam-a}{eq:Gam-c}. In these, $a_{n\kappa}$ and  $c_{n\kappa}$ have to be expressed in terms of $\varphi^\pm_{n\kappa}$ through \Eq{eq:ac-inv}. $\Gamma^u_{n\kappa}$ remains unchanged as in \Eq{eq:Gam-u}. 
Then, $\{\Gamma^+_{n\kappa}, \Gamma^-_{n\kappa}, \Gamma^u_{n\kappa}\}$ are 
quadratic functionals of $\{\varphi^+_{n\kappa}, \varphi^-_{n\kappa}, u_{n\kappa}\}$, 
given by:
\begin{subequations}
\begin{align}
\Gamma^+_{n\kappa} \;=\; \sum_{\ell, m} \,\delta_{\ell m, n}\!\!\int\!
&\left[ 
A^+\varphi^+_{\ell 1}\varphi^+_{m2} \,+\, 
B^+\varphi^+_{\ell 1}\varphi^-_{m2} \,+\,
C^+\varphi^-_{\ell 1}\varphi^-_{m2} \;+\right.
\nonumber\\[-1ex]
&\left.
G^+\varphi^+_{\ell 1}u_{m2} \,+\, 
H^+\varphi^-_{\ell 1}u_{m2} \,+\,
Uu_{\ell 1}u_{m2}
\right]
\delta(12,\bkap)\frac{\dd\bkap_1\dd\bkap_2}{(2\pi)^2}\,,
\label{eq:Gam-pp}\\[1ex]
\Gamma^-_{n\kappa} \;=\; \sum_{\ell, m} \,\delta_{\ell m, n}\!\!\int\!
&\left[ 
A^-\varphi^+_{\ell 1}\varphi^+_{m2} \,+\, 
B^-\varphi^+_{\ell 1}\varphi^-_{m2} \,+\,
C^-\varphi^-_{\ell 1}\varphi^-_{m2} \;+\right.
\nonumber\\[-1ex]
&\left. 
G^-\varphi^+_{\ell 1}u_{m2} \,+\, 
H^-\varphi^-_{\ell 1}u_{m2} \,+\,
Uu_{\ell 1}u_{m2}
\right]
\delta(12,\bkap)\frac{\dd\bkap_1\dd\bkap_2}{(2\pi)^2}\,,
\label{eq:Gam-mm}\\[1ex]
\Gamma^u_{n\kappa} \;=\; \sum_{\ell, m} \,\delta_{\ell m, n}\!\!\int\!
&\left[ 
A^u\varphi^+_{\ell 1}\varphi^+_{m2} \,+\, 
B^u\varphi^+_{\ell 1}\varphi^-_{m2} \,+\,
C^u\varphi^-_{\ell 1}\varphi^-_{m2} \;+\right.
\nonumber\\[-1ex]
&\left.
G^u\varphi^+_{\ell 1}u_{m2} \,+\, 
H^u\varphi^-_{\ell 1}u_{m2} \,+\,
Vu_{\ell 1}u_{m2}
\right]
\delta(12,\bkap)\frac{\dd\bkap_1\dd\bkap_2}{(2\pi)^2}\,.
\label{eq:Gam-uu}
\end{align} 
\end{subequations}
\noindent
Each of $\{\Gamma^\pm_{n\kappa}, \Gamma^u_{n\kappa}\}$ has in its integrand a homogeneous quadratic polynomial in $\{\varphi^+_{n\kappa}, \varphi^-_{n\kappa}, u_{n\kappa}\}$, with all six combinations present. The expressions for $\{\Gamma^\pm_{n\kappa}, \Gamma^u_{n\kappa}\}$ are similar in form, being distinguished only by their interaction 
coefficients. Of these coefficients, $U$ and $V$ have already been defined in \Eq{eq:coeffs-acu}. The other $15$ coefficients are $\{A, B, C, G, H\}$, each of which comes in three flavours $\{+,-,u\}$; these coefficients are combinations of the
$6$ coefficients $\{L, M, P, Q, R, S\}$ defined in \Eq{eq:coeffs-acu}, and are given in \Eqs{eq:coeffs-pm}{eq:coeffs-u}.

The $\Theta_{n\ell}(\kappa)$  are simple functions defined in \Eq{eq:Theta-sat} which are proportional to as yet undetermined constants $\Lambda_{(n-\ell)}^{(\rm s)}$. These constants are given self-consistently in terms of spectra in \Eq{eq:lamb-spec}; see \Sec{sec:grow} for explicit formulas for the case of growing mode turbulence.
Formally, $\Theta_{n\ell}(\kappa)$ is second order in the fields $\varphi^\pm$, which makes the $\Theta_{n\ell}(\kappa)$ terms in \Eqs{eq:phip-evol}{eq:phim-evol} of third order. However, these can be treated as terms that are linear in $\varphi^\pm$ because 
$\Theta_{n\ell}(\kappa)$ have known functional forms with undetermined strengths 
$\Lambda_{(n-\ell)}^{(\rm s)}$.

The equations preserve the condition, $\bra{\varphi^\pm_{n\kappa}} = \bra{u_{n\kappa}} = 0$, that $\{\varphi^\pm_{n\kappa}, u_{n\kappa}\}$ are zero-mean fields. We note a general property that mode amplitudes inherit from \Eq{eq:acu-symm}. Since $\beta\geq 0$, both 
$\lambda^\pm_{n\kappa}$ are real and even functions of $n$ and 
$\bkap$, so we have: 
\begin{align}
\varphi^\pm_n\!\left(\bkap, t\right) &\;=\; \left[\varphi^\pm_n\!\left(-\bkap, t\right)\right]^* \;=\; -\varphi^\pm_{-n}\!\left(\bkap, t\right)\,,
\nonumber\\
u_n\!\left(\bkap, t\right) &\;=\; u^*_n\!\left(-\bkap, t\right) \;=\; u_{-n}\!\left(\bkap, t\right)\,,\label{eq:pmu-symm}\\
\varphi^\pm_0\!\left(\bkap, t\right) &\,=\, 0\,.
\nonumber
\end{align}
\noindent
Note that $u_0\!\left(\bkap, t\right)$ need not be zero, corresponding to a horizontal mode with no $z$-variation. These properties of $\{\varphi^\pm_{n\kappa}, u_{n\kappa}\}$ are used repeatedly in calculations. 

\subsection{Energy Pathways}\label{sec:en-path}
The system defined by \Eqss{eq:phip-evol}{eq:Gam-uu} is in the desired form for a modal description of nonlinear interactions among the three modes $\{\varphi^\pm_{n\kappa}, u_{n\kappa}\}$. The linear terms give the three linear growth rates, of which those associated with $\varphi^-_{n\kappa}$ and $u_{n\kappa}$ are always negative. All three modes decay at high wavenumbers. We recall from \Eq{grow-crit} that $\gamma^+_{n\kappa} > 0$ whenever $(n, \kappa)$ are such that $\ra > (k_nd)^6/(\kappa d)^2$. For the high $\ra$ convection we consider, there is a range of values of $(n, \kappa)$ for which 
$\gamma^+_{n\kappa} > 0$. Growing gravity waves in this range extract energy from the superadiabatic background. The energy can follow multiple pathways---enabled by the nonlinear interactions in $\{\Gamma^\pm_{n\kappa}, \Gamma^u_{n\kappa}\}$---toward dissipation. 

With $18$ coefficients ($U$ is repeated twice) determining these nonlinear interactions, the variety in pathways is an embarrassment of riches. The relative efficiency of energy flow among the many possible pathways is a question of much interest. The problem would be considerably simplified if  a few pathways were much more efficient than many others. Indeed, physical reasoning has favoured the notion of a dominant pathway, which is explored below.

\subsubsection{Growing Mode Turbulence}\label{sec:grow}
The traditionally dominant pathway \citep{lss61,yam63} involves the turbulent cascade of the growing modes, while the dynamics of decaying modes and horizontal modes remains subordinate. This system can be precisely described by drastically reducing the nonlinear terms in \Eqss{eq:phip-evol}{eq:u-evol-new}. In this model, the assumption $\vert|\varphi^+\vert| \gg \vert|\varphi^-\vert|, \vert|u\vert|$ is proposed as an \emph{ansatz}. Setting $\varphi^- = u = 0$ in the second- and third-order terms of \Eq{eq:phip-evol} simplifies this equation for the growing mode to:
\begin{align}
\frac{\partial \varphi^+_{n\kappa}}{\partial t} \;=\;  
\gamma^+_{n\kappa}\varphi^+_{n\kappa} 
&\;+\; \sum_{\ell, m} \,\delta_{\ell m, n}\!\!
\int\! A^+\!\!\left[\,\varphi^+_{\ell 1}\varphi^+_{m2} \,-\, \bra{\varphi^+_{\ell 1}\varphi^+_{m2}}\,\right]\delta(12,\bkap)\frac{\dd\bkap_1\dd\bkap_2}{(2\pi)^2}
\nonumber\\
&\;+\; \lambda^+_{n\kappa}\sum_{\ell}{\cal T}_{n\ell}(\kappa)\,\varphi^+_{\ell\kappa}\,,
\qquad\mbox{for $n=1,2,\ldots\,$,}
\label{eq:phip-dom}
\end{align}
\noindent
where $A^+$ are the sole surviving interaction coefficients, defined in \Eq{eq:coeffs-pm}, characterizing mutual quadratic interactions between growing modes. The coefficients,
\beq
{\cal T}_{n\ell}(\kappa) \;=\; 
\begin{cases} 
   \; 0 & \quad \text{if}\quad \ell = n,\\[1ex]
   \; \displaystyle \frac{\Lambda_{(n-\ell)}^{(\rm s)}}{\chi}\frac{\lambda^-_{\ell\kappa}}  
   {\left(\lambda^+_{\ell\kappa} - \lambda^-_{\ell\kappa}\right)}\frac{\kappa}{k_\ell}
        & \quad \text{otherwise,}
\end{cases}
\label{eq:T-sat}
\eeq
are well-defined functions of $\ell$ and $\kappa$ multiplied by as-yet-undetermined constants, $\Lambda_{(n-\ell)}^{(\rm s)}$. The ${\cal T}_{n\ell}(\kappa)$ terms in \Eq{eq:phip-dom} couple the saturated mean temperature profile and the growing mode amplitudes. These terms can be treated in a formal manner as linear couplings to all 
$\varphi^+_{\ell\kappa}$, with $\ell\neq n$. The system becomes self-consistent when the coupling strengths $\Lambda_{(n-\ell)}^{(\rm s)}$ are expressed in terms of the saturated spectra of the $\varphi^+$ fields; this is done below. Using the definition in \Eq{eq:lamb-spec}, we have: 
\beq
\Lambda_m^{(\rm s)} \;=\; \sum_{i, j}\delta_{i j, m}\!\!\int\frac{\kappa}{k_j} 
\frac{(-\lambda^-_{j\kappa})}  
   {\left(\lambda^+_{i\kappa} - \lambda^-_{i\kappa}\right)\left(\lambda^+_{j\kappa} - \lambda^-_{j\kappa}\right)} {\cal P}^{++}_{ij}(\kappa)\,\kappa\,\dd\kappa\,,
\label{eq:lamb-pp}
\eeq
where ${\cal P}^{++}_{ij}(\kappa)$ are saturated spectra of the growing modes, defined in terms of saturated two-point correlators by: 
\beq
\bra{\varphi^+_i(\bkap)\varphi^{+*}_j(\bkap')}^{(\rm s)} \;=\; (2\pi)^3\, {\cal P}^{++}_{ij}(\kappa)\,\delta(\bkap - \bkap')\,.
\label{eq:spec-pp}
\eeq

\Eq{eq:phip-dom} is the amplitude equation governing growing mode turbulence. The first term on the right-hand side represents linear growth. As discussed at the beginning of \Sec{sec:en-path}, in high $\ra$ convection, $\gamma^+_{n\kappa} > 0$ for a range of $(n, \kappa)$. Starting from very small values, fluctuations in the $\varphi^+$ lying in this range extract energy from the superadiabatic background. The second term describes mutual quadratic interactions---mediated by the coefficients $A^+$---among the $\varphi^+$, transfering energy to modes lying inside and outside this range. The third term with the 
${\cal T}$ describes the effect of the saturated mean temperature profile; it spreads the energy among the $\varphi^+$ through effectively linear couplings, whose strengths are to be determined self-consistently. At large enough $(n, \kappa)$, the growth rates 
$\gamma^+_{n\kappa}$ become negative because  of thermal and viscous dissipation; hence, the turbulent cascade dissipates. Of primary interest is the determination of the saturated growing mode spectra, ${\cal P}^{++}_{ij}(\kappa)$.

\subsubsection{Excitation of Decaying Modes and Horizontal Modes}\label{sec:dec-u}
Continuing with the \emph{ansatz} $\vert|\varphi^+\vert| \gg \vert|\varphi^-\vert|, \vert|u\vert|$, we again set $\varphi^- = u = 0$ in all the nonlinear terms of    
\Eqs{eq:phim-evol}{eq:u-evol-new}. This gives:
\begin{subequations}
\begin{align}
\frac{\partial \varphi^-_{n\kappa}}{\partial t} \;=\;  
\gamma^-_{n\kappa}\varphi^-_{n\kappa} 
&\;+\; \sum_{\ell, m} \,\delta_{\ell m, n}\!\!
\int\! A^-\!\!\left[\,\varphi^+_{\ell 1}\varphi^+_{m2} \,-\, \bra{\varphi^+_{\ell 1}\varphi^+_{m2}}\,\right]\delta(12,\bkap)\frac{\dd\bkap_1\dd\bkap_2}{(2\pi)^2}
\nonumber\\
&\;+\; \lambda^-_{n\kappa}\sum_{\ell}{\cal T}_{n\ell}(\kappa)\,\varphi^+_{\ell\kappa}\,,
\qquad\mbox{for $n=1,2,\ldots\,$,}
\label{eq:phim-dom}\\[1em]
\frac{\partial u_{n\kappa}}{\partial t} \;=\;  
-\nu k_n^2 u_{n\kappa}
&\;+\; \sum_{\ell, m} \,\delta_{\ell m, n}\!\!
\int\! A^u\!\!\left[\,\varphi^+_{\ell 1}\varphi^+_{m2} \,-\, \bra{\varphi^+_{\ell 1}\varphi^+_{m2}}\,\right]\delta(12,\bkap)\frac{\dd\bkap_1\dd\bkap_2}{(2\pi)^2}\,,
\nonumber\\
&\mbox{for $n=0,1,2,\ldots\,$,}
\label{eq:u-dom}
\end{align}
\end{subequations}
\noindent
where the interaction coefficients $A^-$ and $A^u$ are defined in \Eqs{eq:coeffs-pm}{eq:coeffs-u}.

These equations describe the time evolution of $\varphi^-_{n\kappa}$ and $u_{n\kappa}$ when they are excited by the dominant $\varphi^+$ modes and dissipate because the linear growth rates, $\gamma^-_{n\kappa}$ and $-\nu k_n^2$, are always negative. The main source of excitation of these subordinate modes are the quadratic terms with interaction coefficients $A^-$ and $A^u$; in addition, $\varphi^-_{n\kappa}$ is also linearly excited by the $\varphi^+_{\ell\kappa}$ modes, with coupling coefficients ${\cal T}_{n\ell}(\kappa)$ related to the saturated mean temperature profile. The power- and cross-spectra of the $\varphi^-$ and $u$ modes will be entirely determined by those of the growing modes. \Eqss{eq:phip-dom}{eq:u-dom} provide a precise model to enable computation of these saturated spectra.

\section{Conclusions}\label{sec:con}
The modal equations are the end result of a progression in the symmetry and uniformity of the variables used to describe the fully developed turbulent state of Boussinesq convection, in which statistical homogeneity and isotropy hold in the horizontal directions. \Eqss{eq:phip-evol}{eq:Gam-uu} formulate turbulent convection in terms of  
coupled nonlinear interactions among three mode types: growing gravity modes $\varphi^+$, decaying gravity modes $\varphi^-$, and horizontal modes $u$. These equations govern the 
time evolution of the zero-mean fluctuations $\{\varphi^\pm, u\}$, describing not only the saturated turbulence explored in the numerical simulations of \Secs{sec:num1}{sec:num2} but also the approach to saturation. The mean temperature profile is determined self-consistently along with the power- and cross-spectra of $\{\varphi^\pm, u\}$. The energy input from the buoyancy forces, through the excitation of $\varphi^+$, can follow multiple pathways toward dissipation because all possible quadratic interactions among the three mode types are allowed.

Numerical explorations are necessary to guide us in the search for dominant pathways. The simulation of \Secs{sec:num1}{sec:num2} can be analyzed to test the assumption, 
$\vert|\varphi^+\vert| \gg \vert|\varphi^-\vert|, \vert|u\vert|$, underlying the traditionally considered pathway \citep{lss61,yam63}, in which convective turbulence is dominated by the $\varphi^+$ modes, with $\varphi^-$ and $u$ playing subordinate roles. Proposing this assumption as an \emph{ansatz}, in \Secs{sec:grow}{sec:dec-u} we derived a model in which the modal equations simplify considerably. \Eq{eq:phip-dom} governs the time evolution of the dominant $\varphi^+$, growth due to linear instability, nonlinear interactions among the $\varphi^+$ themselves, and linear dissipation at high wavenumbers. These equations for the $\varphi^+$ underly growing mode turbulence; $\varphi^-$ and $u$ do not occur in them. The growing mode spectra, ${\cal P}^{++}(\kappa)$, contain all the information needed to calculate first approximations to important physical quantities like heat fluxes and energies. These spectra can be calculated from kinetic models based on \Eq{eq:phip-dom}. The next level of approximation considers the nonlinear excitation of $\varphi^-$ and $u$ modes by the $\varphi^+$, and their linear dissipation, as described by \Eqs{eq:phim-dom}{eq:u-dom}. The various modal power- and cross-spectra involving $\varphi^-$ and $u$ are entirely controlled by the ${\cal P}^{++}(\kappa)$. 

While large-scale modes predominantly govern convective heat transport, understanding the energy cascade to high-wavenumber regimes ($n$ and $\kappa$) is crucial for a complete physical picture. Taking the limit of large $n$, we can treat the vertical wavenumber $q_n = n\pi/d \to k_z$ as continuous. This continuous treatment significantly simplifies the modal equations, enabling the use of standard three-dimensional Fourier transforms rather than a cumbersome combination of Fourier series (in $z$) and two-dimensional transforms (in the horizontal plane). These modal equations provide a robust theoretical foundation for kinetic models, allowing us to explore small-scale turbulence structures and the anisotropy of energy flow in $\kk \equiv (\bkap, k_z)$ space.

We conclude with comments on two possible extensions of the present work. Although we have focused on the case of convectively unstable stratification, our formalism also applies---with minor modifications---to stably stratified atmospheres. In this case, where $\beta < 0$, the corresponding modal equations would describe the nonlinear interactions among internal gravity waves and horizontal modes. Adopting an approach similar to that followed here, rotating RBC formulations \citep{bdl14,es23} can also be cast in terms of modal equations; however, this would entail rather more labor, in accordance with the more complex physical situation.

\appendix
\section{Criterion for Growing Gravity Modes}\label{app:ra-crit}
There is a growing gravity mode with wavevector $\kk_n$ when $\gamma_n^{+}(\kappa)$ (as defined in \Eq{gamma-pm}) is positive. When either $\nu$ or $\chi$ is zero, this implies 
that the superadiabatic gradient, $\beta$, must be positive. For the generic case when neither $\nu$ nor $\chi$ is zero, a growing mode implies that:   
\beq
\ra \;>\; \frac{(k_n d)^6}{(\kappa d)^2} = \frac{\left[(n\pi)^2 + (\kappa d)^2\right]^3}{(\kappa d)^2}\,.
\label{grow-crit}
\eeq
\noindent
The critical Rayleigh number, $\ra_{\rm cr}$, is the minimum value of the right-hand side of \Eq{grow-crit}, and the mode (or modes) for which this value is attained may be referred to as the critical mode (or modes). For a given $\kappa d$, the right-hand side of \Eq{grow-crit} is smallest when $n^2$ takes its lowest value. Since $n\neq 0$ for gravity modes, the lowest value of $n^2$ is unity.

With $n^2 =1$, the right-hand side of \Eq{grow-crit} assumes its minimum value when 
$(\kappa d)^2 = \pi^2/2$ for the case of infinite lateral boundary conditions. This gives
$\ra_{\rm cr} = (27/4)\pi^4 = 657.51\ldots\,$ and a critical mode with $n=\pm 1$ and horizontal wavelength $\sqrt{8}d$. For periodic lateral boundary conditions applied to a square box of size $L\times L\times d$, $\kappa_x = \ell(2\pi/L)$, $\kappa_y = m(2\pi/L)$, with $\ell$ and $m$ integers. The right-hand side of \Eq{grow-crit} is smallest when $n^2=1$, and $(\kappa d)^2 = (2\pi d/L)^2\left[\ell^2 + m^2\right]$ is close to 
$\pi^2/2$. For the numerical simulation discussed in \Sec{sec:num1}, where $L/d = 8$, it turns out $(\kappa d)^2$ is exactly equal to $\pi^2/2$ when $\ell=\pm 2$ and $m = \pm 2$. Therefore, we again have $\ra_{\rm cr} = 657.51\ldots\,$ and a critical mode with horizontal wavelength $\sqrt{8}d$. Since the numerical simulation used $\ra = 10^6$, we expect many growing modes to be excited in addition to the critical mode. 

\section{Interaction Coefficients}\label{app:int-coeff}
The eight coefficients that occur in the expressions for $\{\Gamma^a_{n\kappa}, \Gamma^c_{n\kappa}, \Gamma^u_{n\kappa}\}$, defined in \Eqss{eq:Gam-a}{eq:Gam-u}, are:
\begin{align}
L &\;\equiv\; L_{n\ell m}(\bkap, \bkap_1, \bkap_2) \;=\; 
(\ea_{n\kappa}\cendot\,\ea_{\ell 1})(\kk_n\cendot\,\ea_{m2})\,,
\nonumber\\[1ex]
M &\;\equiv\; M_{n\ell}(\bkap, \bkap_1, \bkap_2) \;=\; 
(\ea_{n\kappa}\cendot\,\ea_{\ell 1})(\kk_n\cendot\,\eu_{2}) 
\;+\; (\ea_{n\kappa}\cendot\,\eu_{2})(\kk_n\cendot\,\ea_{\ell 1})\,,
\nonumber\\[1ex]
U &\;\equiv\; U_{n}(\bkap, \bkap_1, \bkap_2) \;=\; 
(\ea_{n\kappa}\cendot\,\eu_{1})(\kk_n\cendot\,\eu_{2})\,,
\nonumber\\[1ex]
P &\;\equiv\; P_{nm}(\bkap,\bkap_2) \;=\; (\kk_n\cendot\,\ea_{m2})\,,
\quad
Q \;\equiv\; Q_{n}(\bkap,\bkap_2) \;=\; (\kk_n\cendot\,\eu_{2})\,,
\label{eq:coeffs-acu}\\[1ex]
R &\;\equiv\; R_{n\ell m}(\bkap, \bkap_1, \bkap_2) \;=\; 
(\eu_{\kappa}\cendot\,\ea_{\ell 1})(\kk_n\cendot\,\ea_{m2})\,,
\nonumber\\[1ex]
S &\;\equiv\; S_{n\ell}(\bkap, \bkap_1, \bkap_2) \;=\; 
(\eu_{\kappa}\cendot\,\ea_{\ell 1})(\kk_n\cendot\,\eu_{2}) 
\;+\; (\eu_{\kappa}\cendot\,\eu_{2})(\kk_n\cendot\,\ea_{\ell 1})\,,
\nonumber\\[1ex]
V &\;\equiv\; V_{n}(\bkap, \bkap_1, \bkap_2) \;=\; 
(\eu_{\kappa}\cendot\,\eu_{1})(\kk_n\cendot\,\eu_{2})\,.
\nonumber
\end{align}

The ten coefficients that occur in the expressions for $\Gamma^\pm_{n\kappa}$, defined in \Eqs{eq:Gam-pp}{eq:Gam-mm}, are:
\begin{align}
A^\pm \;\equiv\; A^\pm_{n\ell m}(\bkap, \bkap_1, \bkap_2) \;=\;& 
\lambda^-_{m2}\,
\frac{\lambda^-_{\ell 1}L_{n\ell m}(\bkap, \bkap_1, \bkap_2)  \,-\,   \lambda^\pm_{n\kappa}P_{nm}(\bkap,\bkap_2)}  
{\left(\lambda^+_{\ell 1} - \lambda^-_{\ell 1}\right)\left(\lambda^+_{m2} - \lambda^-_{m2}\right)}\,,
\nonumber\\[3ex]
B^\pm \;\equiv\; B^\pm_{n\ell m}(\bkap, \bkap_1, \bkap_2) \;=\;& 
\lambda^+_{m2}\,
\frac{\lambda^\pm_{n\kappa}P_{nm}(\bkap,\bkap_2) \,-\, 
\lambda^-_{\ell 1}L_{n\ell m}(\bkap, \bkap_1, \bkap_2)}  
{\left(\lambda^+_{\ell 1} - \lambda^-_{\ell 1}\right)\left(\lambda^+_{m2} - \lambda^-_{m2}\right)}
\nonumber\\[1ex]
&\;+\; \lambda^-_{\ell 1}\,
\frac{\lambda^\pm_{n\kappa}P_{n\ell}(\bkap,\bkap_1) \,-\, 
\lambda^+_{m2}L_{nm\ell}(\bkap, \bkap_2, \bkap_1)}  
{\left(\lambda^+_{\ell 1} - \lambda^-_{\ell 1}\right)\left(\lambda^+_{m2} - \lambda^-_{m2}\right)}\,,
\nonumber\\[3ex]
C^\pm \;\equiv\; C^\pm_{n\ell m}(\bkap, \bkap_1, \bkap_2) \;=\;& 
\lambda^+_{m2}\,
\frac{\lambda^+_{\ell 1}L_{n\ell m}(\bkap, \bkap_1, \bkap_2)  \,-\,   \lambda^\pm_{n\kappa}P_{nm}(\bkap,\bkap_2)}  
{\left(\lambda^+_{\ell 1} - \lambda^-_{\ell 1}\right)\left(\lambda^+_{m2} - \lambda^-_{m2}\right)}\,,
\nonumber\\[3ex]
G^\pm \;\equiv\; G^\pm_{n\ell}(\bkap, \bkap_1, \bkap_2) \;=\;& 
\frac{\lambda^\pm_{n\kappa}Q_{n}(\bkap,\bkap_2) \,-\, 
\lambda^-_{\ell 1}M_{n\ell}(\bkap, \bkap_1, \bkap_2)}  
{\left(\lambda^+_{\ell 1} - \lambda^-_{\ell 1}\right)}\,,
\nonumber\\[3ex]
H^\pm \;\equiv\; H^\pm_{n\ell}(\bkap, \bkap_1, \bkap_2) \;=\;& 
\frac{\lambda^+_{\ell 1}M_{n\ell}(\bkap, \bkap_1, \bkap_2) \,-\, \lambda^\pm_{n\kappa}Q_{n}(\bkap,\bkap_2)}  
{\left(\lambda^+_{\ell 1} - \lambda^-_{\ell 1}\right)}\,.
\label{eq:coeffs-pm}
\end{align}

The five coefficients that occur in the expression for $\Gamma^u_{n\kappa}$, defined in \Eq{eq:Gam-uu}, are:
\begin{align}
A^u &\;\equiv\; A^u_{n\ell m}(\bkap, \bkap_1, \bkap_2) \;=\; 
\frac{\lambda^-_{\ell 1}\lambda^-_{m2}R_{n\ell m}(\bkap, \bkap_1, \bkap_2)}  
{\left(\lambda^+_{\ell 1} - \lambda^-_{\ell 1}\right)\left(\lambda^+_{m2} - \lambda^-_{m2}\right)}\,,
\nonumber\\[3ex]
B^u &\;\equiv\; B^u_{n\ell m}(\bkap, \bkap_1, \bkap_2) \;=\; 
-2\frac{\lambda^-_{\ell 1}\lambda^+_{m2}R_{n\ell m}(\bkap, \bkap_1, \bkap_2)}  
{\left(\lambda^+_{\ell 1} - \lambda^-_{\ell 1}\right)\left(\lambda^+_{m2} - \lambda^-_{m2}\right)}\,,
\nonumber\\[3ex]
C^u &\;\equiv\; c^u_{n\ell m}(\bkap, \bkap_1, \bkap_2) \;=\; 
\frac{\lambda^+_{\ell 1}\lambda^+_{m2}R_{n\ell m}(\bkap, \bkap_1, \bkap_2)}  
{\left(\lambda^+_{\ell 1} - \lambda^-_{\ell 1}\right)\left(\lambda^+_{m2} - \lambda^-_{m2}\right)}\,,
\nonumber\\[3ex]
G^u &\;\equiv\; G^u_{n\ell}(\bkap, \bkap_1, \bkap_2) \;=\;
-\frac{\lambda^-_{\ell 1}S_{n\ell}(\bkap, \bkap_1, \bkap_2)}  
{\left(\lambda^+_{\ell 1} - \lambda^-_{\ell 1}\right)}\,,
\nonumber\\[3ex]
H^u &\;\equiv\; H^u_{n\ell}(\bkap, \bkap_1, \bkap_2) \;=\; 
\frac{\lambda^+_{\ell 1}S_{n\ell}(\bkap, \bkap_1, \bkap_2)}  
{\left(\lambda^+_{\ell 1} - \lambda^-_{\ell 1}\right)}\,. 
\label{eq:coeffs-u}
\end{align}

\end{document}